\documentclass[acmsmall,nonacm,screen]{acmart}
\setcopyright{none}
\renewcommand\acmJournal[1]{}
\renewcommand\acmVolume[1]{}
\renewcommand\acmNumber[1]{}
\renewcommand\acmArticle[1]{}
\renewcommand\acmYear[1]{}
\renewcommand\acmMonth[1]{}

\AtBeginDocument{%
  }

\usepackage{enumitem}
\usepackage{array}
\usepackage{wrapfig}
\usepackage{amsmath,amsfonts}
\usepackage[noend]{algpseudocode}
\usepackage{graphicx}
\usepackage{textcomp}
\usepackage{float}
\usepackage{listings}
\usepackage{xspace}
\usepackage{multirow}
\usepackage{amsthm}

\usepackage{balance}
\usepackage{algorithm}
\usepackage{algpseudocode}
\usepackage{colortbl}
\usepackage{subcaption}

\usepackage[skins]{tcolorbox}
\usepackage{xcolor}
\usepackage{multicol}
\usepackage{soul}
\usepackage{framed}
\usepackage{booktabs}
\usepackage{booktabs}
\usepackage{multirow}
\usepackage{siunitx}

\usepackage{algorithm}
\usepackage{tabularx}
\usepackage{booktabs}
\usepackage{color}
\usepackage{multirow}
\usepackage{listings}
\usepackage{pythonhighlight}
\usepackage{subcaption}
\usepackage{tikz}
\usetikzlibrary{positioning}
\usepackage{tabularx}
\usepackage{algpseudocode}
\usepackage{graphicx}
\usepackage{balance}
\usepackage{pgfplotstable}
\usepackage{tcolorbox}
\usepackage{lipsum}
\usepackage{enumitem}

\usepackage{makecell}

\usepackage[normalem]{ulem}
\newcommand*\colourcheck[1]{%
	\expandafter\newcommand\csname #1check\endcsname{\textcolor{#1}{\ding{52}}}%
}
\colourcheck{blue}
\colourcheck{green}
\colourcheck{red}
\newtcolorbox{boxB}[2][]{%
  enhanced,colback=white,colframe=black,coltitle=black,
  sharp corners,
  toprule=1.0pt,
  rightrule=0.3pt,
  leftrule=0pt,
  bottomrule=0pt,
  fonttitle=\itshape\scshape\large,
  left=0pt,right=5pt,top=5pt,bottom=3pt,
  attach boxed title to top right={yshift=-0.3\baselineskip-0.4pt,xshift=-5mm},
  boxed title style={tile,size=minimal,left=0.2mm,right=0.5mm,
    colback=white,before upper=\strut},
  title=#2,#1
}

\newtcolorbox{outputbox}[1][]{ 
  colback=lightgray!10,  
  colframe=gray,      
  fontupper=\scriptsize,
  left=1mm, right=1mm, top=0.5mm, bottom=0.5mm,  
  before=\definecolor{royalblue}{rgb}{0.2549, 0.4118, 0.8824} 
}

\definecolor{lightred}{rgb}{1,0.85,0.85}
\definecolor{lightblue}{rgb}{0.85,0.92,1}
\definecolor{lightgreen}{rgb}{0.85,1,0.85}

\newtcolorbox{zoneboxred}{
  colback=lightred,
  colframe=red!50!red!50,
  boxrule=0pt,
  sharp corners,
  enhanced,
  left=1mm, right=1mm, top=0.5mm, bottom=0.5mm,
  before skip=1mm, after skip=1mm
}

\newtcolorbox{zoneboxblue}{
  colback=lightblue,
  colframe=blue!50!blue!50,
  boxrule=0pt,
  sharp corners,
  enhanced,
  left=1mm, right=1mm, top=0.5mm, bottom=0.5mm,
  before skip=1mm, after skip=1mm
}

\newtcolorbox{zoneboxgreen}{
  colback=lightgreen,
  colframe=green!50!green!50,B
  boxrule=0pt,
  sharp corners,
  enhanced,
  left=1mm, right=1mm, top=0.5mm, bottom=0.5mm,
  before skip=1mm, after skip=1mm
}

\newboolean{showcomments}
\setboolean{showcomments}{true}
\ifthenelse{\boolean{showcomments}}
 { \newcommand{\mynote}[2]{
      \fbox{\bfseries\sffamily\scriptsize#1}
        {\small$\blacktriangleright$\textsf{\emph{#2}}$\blacktriangleleft$}}}
        { \newcommand{\mynote}[2]{}}

\usepackage{tikz}

\newcolumntype{L}[1]{>{\raggedright\arraybackslash}p{#1}}

\newcommand{\code}[1]{{\footnotesize\texttt{#1}}}
\usepackage{amsthm}
\definecolor{dkgreen}{rgb}{0,0.6,0}
\definecolor{gray}{rgb}{0.5,0.5,0.5}
\definecolor{lightgray}{rgb}{211, 211, 211}
\definecolor{mauve}{rgb}{0.58,0,0.82}

\definecolor{custom-red}{rgb}{1,0,0}
\definecolor{custom-blue}{rgb}{0,0,1}

\definecolor{c5}{HTML}{ffffff}
\definecolor{c6}{HTML}{fffdfd}
\definecolor{c7}{HTML}{f5cfcf}
\definecolor{c8}{HTML}{fffbfb}
\definecolor{c9}{HTML}{ffffff}
\definecolor{c10}{HTML}{fffdfd}
\definecolor{c11}{HTML}{fefafa}
\definecolor{c12}{HTML}{fef7f7}
\definecolor{c13}{HTML}{ffffff}
\definecolor{c14}{HTML}{fffefe}
\definecolor{c15}{HTML}{ffffff}
\definecolor{c16}{HTML}{fefafa}
\definecolor{c17}{HTML}{fdf3f3}
\definecolor{c18}{HTML}{fffefe}
\definecolor{c19}{HTML}{fdf5f5}
\definecolor{c20}{HTML}{ffffff}

\definecolor{customcolor}{HTML}{CAEEFB}

\definecolor{dkgreen}{rgb}{0,0.6,0}
\definecolor{gray}{rgb}{0.5,0.5,0.5}
\definecolor{mauve}{rgb}{0.58,0,0.82}
\lstdefinelanguage{JavaScript}{
  keywords={
    break, case, catch, class, const, continue, debugger, default,
    delete, do, else, export, extends, finally, for, function, if,
    import, in, instanceof, let, new, return, super, switch, this,
    throw, try, typeof, var, void, while, with, yield
  },
  sensitive=true,
  comment=[l]{//},
  comment=[s]{/*}{*/},
  string=[b]",
}

\usepackage{tcolorbox}
\tcbuselibrary{listings,skins,breakable}
\usepackage{subcaption}
\usepackage{tikz}
\usetikzlibrary{arrows.meta,positioning}

\usepackage{listings}
\lstdefinestyle{compact}{
  basicstyle=\ttfamily\scriptsize,
  columns=fullflexible,
  keepspaces=true,
  showstringspaces=false,
  breaklines=true,
  breakatwhitespace=true,
  xleftmargin=0.4em,
  frame=none
}

\newtcblisting{codepanel}[3]{%
  enhanced,
  breakable,
  colback=white,
  colframe=#2,
  title=\textbf{#1},
  coltitle=white,
  colbacktitle=#2,
  fonttitle=\tiny,
  attach boxed title to top center={yshift=-1mm},
  boxed title style={boxrule=0pt, arc=8pt},
  listing engine=listings,
  listing options={
    language=#3,
    basicstyle=\tiny\ttfamily,
    columns=fullflexible,
    keepspaces=true,
    breaklines=true,
    breakatwhitespace=false,
    showstringspaces=false
  }
}

\lstdefinestyle{textStyle}{
    keywordstyle= \color{blue!70},
    commentstyle= \color{red!50!green!50!blue!50},
    stringstyle=\color{red!70},
    frame=shadowbox,
    rulesepcolor= \color{red!20!green!20!blue!20} ,
    xleftmargin=1.5em,xrightmargin=0em, aboveskip=1em,
    framexleftmargin=1.5em,
            numbersep= 5pt,
    basicstyle=\scriptsize\ttfamily,
    numberstyle=\scriptsize\ttfamily,
    emphstyle=\bfseries,
    numbers=none
}

\definecolor{codegray}{gray}{0.5}
\definecolor{codepurple}{rgb}{0.58,0,0.82}
\definecolor{backcolour}{rgb}{0.95,0.95,0.92}
\definecolor{blue}{rgb}{0,0,1}

\lstdefinestyle{pythonStyle}{
    backgroundcolor=\color{backcolour},
    commentstyle=\color{codegray},
    keywordstyle=\color{magenta},
    numberstyle=\tiny\color{codegray},
    stringstyle=\color{codepurple},
    basicstyle=\ttfamily\footnotesize,
    breakatwhitespace=false,
    breaklines=true,
    captionpos=b,
    keepspaces=true,
    numbers=left,
    numbersep=5pt,
    showspaces=false,
    showstringspaces=false,
    showtabs=false,
    tabsize=2,
    language=Python,
    escapeinside={(*@}{@*)}
}

\lstdefinestyle{compact}{
  basicstyle=\footnotesize\ttfamily,
  numbers=left,
  numberstyle=\scriptsize\color{gray},
  numbersep=3pt,
  backgroundcolor=\color{lightgray},
  xleftmargin=0.9em,
  framexleftmargin=0.9em,
  xrightmargin=0em,
  aboveskip=0.35em,
  belowskip=0.35em,
  breaklines=true,
  breakautoindent=false,
  showstringspaces=false,
  columns=fullflexible
}

\lstdefinestyle{htmlStyle}{
    language=Python,
    numbers=left,
    numberstyle= \tiny,
    keywordstyle= \color{blue!70},
    commentstyle= \color{red!50!green!50!blue!50},
    stringstyle=\color{red!70},
    frame=shadowbox,
    rulesepcolor= \color{red!20!green!20!blue!20} ,
    xleftmargin=1.5em,xrightmargin=0em, aboveskip=1em,
    framexleftmargin=1.5em,
            numbersep= 5pt,
    basicstyle=\scriptsize\ttfamily,
    numberstyle=\scriptsize\ttfamily,
    emphstyle=\bfseries,
    morekeywords={doctype, html, head, body, title, script, div, span, a, data-result} 
}

\definecolor{eclipseStrings}{RGB}{42,0.0,255}
\definecolor{eclipseKeywords}{RGB}{127,0,85}
\colorlet{numb}{magenta!60!black}

\lstdefinelanguage{json}{
    basicstyle=\scriptsize\ttfamily,
    numberstyle=\scriptsize\ttfamily,
    emphstyle=\bfseries,
    commentstyle=\color{eclipseStrings}, 
    stringstyle=\color{eclipseKeywords}, 
    numbers=left,
    numberstyle= \tiny,
    keywordstyle= \color{blue!70},
    commentstyle= \color{red!50!green!50!blue!50},
    frame=shadowbox,
    rulesepcolor= \color{red!20!green!20!blue!20} ,
    xleftmargin=1.5em,xrightmargin=0em, aboveskip=1em,
    framexleftmargin=1.5em,
            numbersep= 5pt,
    showstringspaces=false,
    breaklines=true,
    string=[s]{"}{"},
    comment=[l]{:\ "},
    morecomment=[l]{:"},
    literate=
        *{0}{{{\color{numb}0}}}{1}
         {1}{{{\color{numb}1}}}{1}
         {2}{{{\color{numb}2}}}{1}
         {3}{{{\color{numb}3}}}{1}
         {4}{{{\color{numb}4}}}{1}
         {5}{{{\color{numb}5}}}{1}
         {6}{{{\color{numb}6}}}{1}
         {7}{{{\color{numb}7}}}{1}
         {8}{{{\color{numb}8}}}{1}
         {9}{{{\color{numb}9}}}{1}
}

\usepackage{tikz}

\definecolor{request}{HTML}{D80073}
\definecolor{response}{HTML}{0050EF}
\definecolor{circle_color}{HTML}{CCE5FF}

\usepackage{graphicx}
\usepackage{wrapfig}

\begin{document}

\title[The Effect of Code Obfuscation on Human Program Comprehension]{The Effect of Code Obfuscation on Human Program Comprehension}

\author{Anh H. N. Nguyen}
\authornote{Both authors contributed equally to this research.}
\email{ahn210003@utdallas.edu}
\author{Jack Le}
\authornotemark[1]
\email{jvl210002@utdallas.edu}
\affiliation{%
 \institution{University of Texas at Dallas}
 \city{Richardson}
 \state{Texas}
 \country{USA}
}

\author{Ilse Lahnstein Coronado}
\affiliation{%
 \institution{University of Texas at Dallas}
 \city{Richardson}
 \state{Texas}
 \country{USA}}
\email{ial200001@utdallas.edu}

\author{Tien N. Nguyen}
\affiliation{%
 \institution{University of Texas at Dallas}
 \city{Richardson}
  \state{Texas}
 \country{USA}}
\email{Tien.N.Nguyen@utdallas.edu}

\begin{abstract}
We investigate how code obfuscation influences human understanding of programs through an output-prediction task. To study this effect, we construct multiple levels of obfuscation, ranging from unobfuscated code to transformations involving identifier renaming, adversarially misleading identifiers, control-flow modifications, and combinations of these techniques. These transformations are applied to function-level programs written in Python and JavaScript. Participants were asked to predict program outputs while we recorded correctness, response time, and self-reported programming experience.

Our results show that obfuscation generally increases the time required to reason about code and tends to reduce prediction accuracy. However, the relationship between obfuscation strength and performance is not strictly monotonic and varies across programming languages. JavaScript exhibits the expected pattern of increasing difficulty with stronger obfuscation, whereas Python displays a more complex trend in which certain renaming transformations can perform comparably to, or occasionally better than, the unobfuscated baseline. Response-time analyses further suggest that obfuscation shifts participants away from rapid, heuristic reasoning toward slower and more deliberate reasoning processes. Performance appears highest within a moderate range of response times, indicating that careful deliberation can improve accuracy, while extremely long response times often correspond to confusion. Finally, programming experience predicts performance primarily within a given language, with limited transfer across languages, suggesting that obfuscation challenges language-specific familiarity more than general programming ability.
\end{abstract}

\newcommand{\red}[1]{\textcolor{red}{#1}} 
\newcommand{\blue}[1]{\textcolor{blue}{#1}} 
\newcommand{\magenta}[1]{\textcolor{magenta}{#1}} 

\maketitle

\section{Introduction}
\label{sec:intro}

Software obfuscation is widely used to protect intellectual property, deter reverse engineering, and hide the business logic. In practice, obfuscators routinely transform identifier names and control flow, and they often stack multiple transformations to maximize resistance. Yet the effectiveness of these techniques is usually evaluated through automated metrics (e.g., code complexity, decompilation success, or similarity scores) or against attacker models that implicitly assume how a human analyst will struggle. This leaves a basic gap: {\em obfuscation is fundamentally meant to burden human comprehension}, but we have limited empirical evidence about how specific obfuscation mechanisms change people's ability to understand code.

A common assumption in software security is that stronger obfuscation monotonically increases cognitive difficulty. However, this assumption is not trivially true for humans, as human cognition is not simply computational. Identifier renaming might remove semantic ``shortcuts,'' forcing more careful reasoning; adversarial renaming might mislead by injecting plausible-but-wrong semantics; and control-flow alteration can disrupt the structural cues that programmers rely on to form a mental  model. These obfuscations target different aspects of comprehension, and their effects may not simply add up when combined. Without controlled human studies, it is unclear whether ``more obfuscation'' always means ``harder,'' which techniques actually drive the difficulty, and whether some transformations inadvertently help by pushing readers away from misleading heuristics.

To ground these questions in measurable behavior, we study {\em output prediction}--given a function and a concrete input, determine the exact output. Output prediction is a classic proxy for program comprehension because it requires building a correct mental model of the program's semantics, yet it avoids the ambiguity of open-ended tasks like summarization. It also provides a clean, objective signal (correct/incorrect) and supports fine-grained analysis of response time. In cognitive terms, this lets us examine when programmers rely on fast, heuristic processing versus when they switch to slower, deliberate tracing--an instance of the Dual-System (System 1/System 2)~\cite{evans2013dual} view of~reasoning. Output prediction was also used in prior studies on code understanding~\cite{siegmund2014understanding,siegmund2017measuring}. 

Motivated by these gaps, we conduct a controlled experiment that varies obfuscation along a small, representative hierarchy: L0 (original code), L1 (identifier renaming with uninformative names), L1b (adversarial renaming with plausible-but-misleading names), L2 (control-flow alteration/flattening), and L3 (a realistic combination of renaming and control-flow obfuscation). This design allows us to isolate how different transformations affect both accuracy and time-to-completion, and to test whether the presumed monotonic relationship between obfuscation strength and human difficulty actually holds~\cite{10.1145/2886012} and whether the Dual-System can serve as the grounded theory for such behaviors of humans on obfuscated code. It also enables analysis of how code properties (e.g., cyclomatic complexity, length, identifier length) correlate with performance and cognitive effort, and whether these relationships differ across Python and JavaScript.

Finally, obfuscation is deployed across ecosystems, but human comprehension is shaped by language syntax, idioms, and prior experience. A transformation that is disruptive in one language may be less harmful--or harmful in a different way--in another. Similarly, experience may not transfer cleanly: being fluent in Python may not make a programmer robust to obfuscated JavaScript, and vice versa. Understanding these cross-language and experience effects is essential for interpreting obfuscation ``strength'' in realistic settings and for designing evaluations that reflect how humans actually reason about code. Together, these motivations lead directly to our research questions on how obfuscation affects output prediction accuracy and time, how complexity relates to accuracy, how results differ by language, and how prior experience moderates resilience to~obfuscation.

\section{Background}

\subsection{Code Obfuscation}

\begin{figure}[t]
    \centering
    \includegraphics[width=1\textwidth]{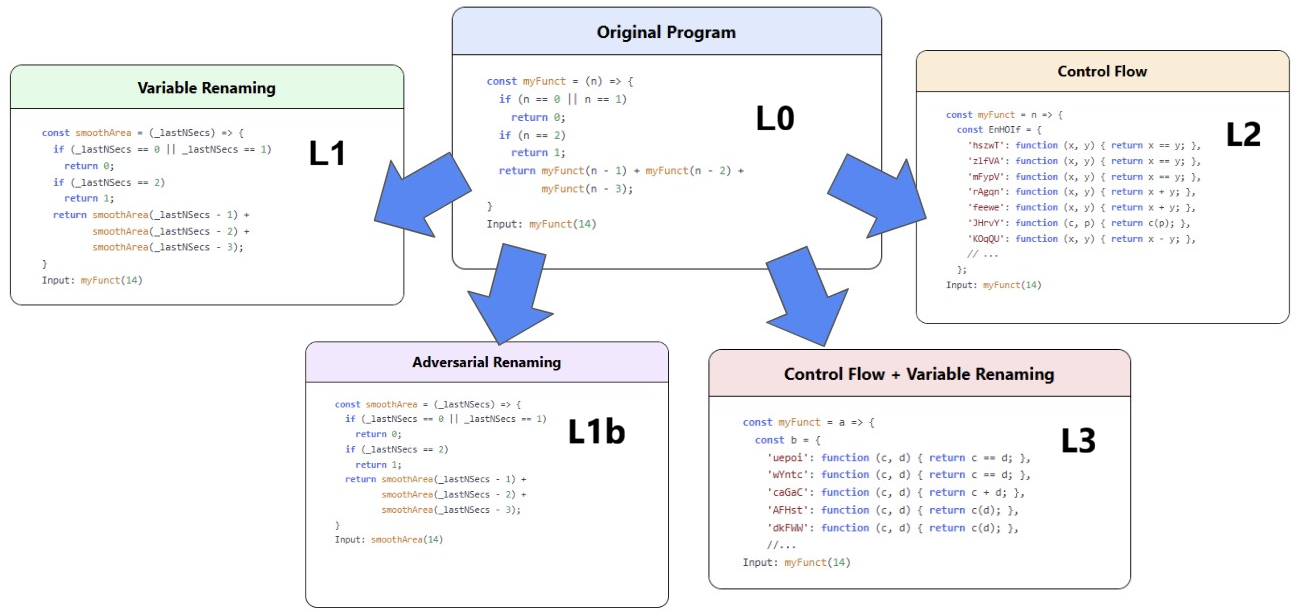}
    \vspace{-18pt}
    \caption{Examples of Obfuscation Levels in Our Study}
    \label{fig:obfuscation-levels}
\end{figure}

Prior work organizes obfuscation techniques by transformation properties, most notably layout-based and control-flow-based obfuscations~\cite{collberg1997taxonomy}. Building on this, we group them in our study into ordinal tiers to enable controlled comparisons of human performance under increasing transformation complexity. This categorization follows the common separation between {\em layout-based} and {\em control flow-based} obfuscation: Identifier Renaming is layout-based, while control-flow flattening is control-flow-based~\cite{ceccato2008towards}. Specifically, our study utilized the following obfuscations (Fig.~\ref{fig:obfuscation-levels}).

\textbf{L1: Identifier Renaming}. Identifier renaming is a layout obfuscation technique that replaces existing identifiers (function names, variable names) into incoherent names, when together, is disjointed semantically. The disconnect cause a harder time understand the program's intent from just the identifiers. 
 In Fig.~\ref{fig:obfuscation-levels}, for the L1 variant, the variables are replaced with short, almost-indistinguishable names, forcing readers to rely on structural reasoning rather than name inference. 

\textbf{L1b: Adversarial Renaming}. We also used L1b, a variant of L1, to capture a different effect of identifier renaming. While L1 reduces readability by renaming identifiers to simplistic, non-informative names, L1b replaces identifiers with semantically meaningful names that are misleading in the new context. This semantic mismatch can mislead readers into forming an incorrect mental model of the code snippet. For example, in Fig.~\ref{fig:obfuscation-levels}, the renamed variable identifier \code{\_lastNSecs} is semantically unrelated to the original identifier \code{n}, and the renamed function \code{smoothArea} is likewise unrelated to the original function name \code{fibfib}.

\textbf{L2: Control-Flow Alteration}. Control-flow alteration obfuscates a program's execution order by decoupling control logic from its original syntactic structure. It typically splits code into smaller blocks and introduces artificial control structures (e.g., dispatchers or indirect jumps) that make the original flow difficult to reconstruct~\cite{laszlo2007cff}. In Fig.~\ref{fig:obfuscation-levels}, the operation (\code{x == y}) is lifted and wrapped in functions stored in an object, so evaluation requires an indirect call (e.g., \code{VuwWxy['mZikr']}) rather than a direct expression. This indirection obscures how execution proceeds.

\textbf{L3: Combination of L1 and L2}. L3 applies both control-flow flattening and variable renaming. Combining multiple obfuscation techniques is common in practice, since developers often layer transformations to better resist reverse engineering. This combination better approximates real-world obfuscated code and allows us to study how the effects of different techniques interact when applied together. By integrating two obfuscation methods, L3 serves as an upper bound on obfuscation difficulty for evaluating the robustness of human code-comprehension performance.

\subsection{Dual System Theory: System 1 and System 2 Thinking}
Cognitive psychology categorizes reasoning into two modes: System 1 (intuitive, fast, heuristic) and System 2 (analytic, slow, deliberate)~\cite{evans2013dual}. The Default-Interventionist Model posits that humans default to low-effort System 1 processing, engaging System 2 only when the default mode fails or conflicts with the task. We hypothesize that obfuscation acts as a barrier to this default mode, necessitating a computationally expensive intervention.

Expert programmers rely on ``beacons,'' familiar features, e.g., variable names and standard structures, to form high-level semantic hypotheses. This enables top-down processing. The Block Model from Schulte~\cite{Schulte2008BlockModel} formalizes this hierarchy, categorizing program comprehension into four levels: (1) Atoms: basic language elements (keywords, variables); (2) Blocks: regions of interest, e.g., a loop body; (3) Relations: dependencies between blocks; and (4) Macro Structure: the overall algorithm or goal.

\begin{wrapfigure}{r}{0.62\textwidth}
\centering
    \includegraphics[width=0.6\textwidth]{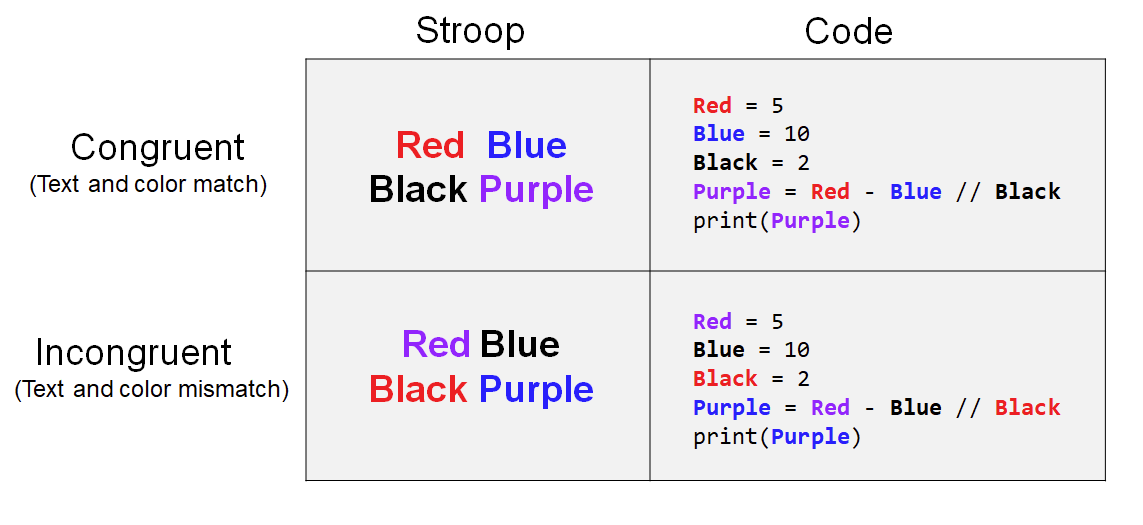}
    \vspace{-18pt}
\    \caption{Stroop and Adversarial Renaming}
    \label{fig:stroop}
\end{wrapfigure}
Obfuscation effectively "flattens" this hierarchy. By removing semantic beacons and altering structure, it obscures the cues necessary for Block and Macro level processing, forcing humans to regress to Atom level analysis, a process that is inherently slower and more error-prone.

Neuroimaging studies validate this dual-process view. Siegmund {\em et al.}~\cite{siegmund2014understanding} studied fMRI signals of programmers on two tasks: output prediction and syntax error correction. 

The authors make the distinction between top-down comprehension is an efficient strategy where programmers rely on prior experience and familiar cues, known as "beacons," to quickly grasp a program's overall purpose without reading every line. 
In contrast, bottom-up comprehension is a lower, statement-by-statement process used for unfamiliar code, requiring the programmer to mentally group individual details into larger "semantic chunks" to build understanding from scratch. In a follow up study, they found that for output prediction, snippets with semantic cues (top-down) required significantly lower brain activation across all relevant areas compared to bottom-up comprehension~\cite{siegmund2017measuring}.

We frame the transition to System 2 as a form of cognitive interference, analogous to the Stroop Interference Effect~\cite{Stroop1935Interference} (see Fig.~\ref{fig:stroop}). Just as incongruent text/color stimuli inhibit automatic reading, obfuscation creates Semantic Incongruence by removing the primes (variable names) that trigger automatic retrieval. The increased response time in our data represents the Inhibition Cost: the latency required to suppress intuitive reading and engage deliberate compilation.

\section{Research Questions}

\textbf{RQ1: How does increasing code obfuscation levels ($L0 \text{--} L3$) affect humans in output prediction?}
Conventional wisdom posits a monotonic relationship between obfuscation levels and cognitive difficulty. 
This RQ seeks to validate the extent to which these assumptions apply to humans.

We choose output prediction as our task due to its importance as a proxy for program comprehension. Unlike summarization, it has a single correct answer and can thus be objectively~measured.

\noindent \textbf{RQ2: How does increasing code obfuscation levels ($L0 \text{--} L3$) affect "time to complete" the output prediction tasks by humans?}

Time-to-completion can be utilized as a proxy for which cognitive mode the user is in: faster responses indicate intuitive heuristics, while slower responses indicate step-by-step, compiler-like processing. Answering this question allows us to verify whether and when participants switch from System 1 to System 2.

\noindent \textbf{RQ3: How does code complexity correlate with human performance?} 
Performance is assessed against objective metrics like cyclomatic complexity and code length. Additionally, we analyze variable name length as a specific stressor on human working memory.

\noindent \textbf{RQ4: How do obfuscation results differ across programming languages (JavaScript vs. Python)?} We analyze how specific language attributes moderate obfuscation. If results differ by programming experience, then obfuscation techniques must be adjusted depending on the programming language. This would mean that the efficacy of obfuscation techniques is dependent on the syntax and rules of that particular language.

\noindent \textbf{RQ5: How does prior programming experience correlate with accuracy in interpreting obfuscated code?} 
Obfuscation can serve as a stress test for the transferability of programming knowledge. Whether or not knowledge in one language helps in comprehending obfuscated code in another language will show the transferability of programming language knowledge in humans.

\section{Experimental Setup}

\subsection{Data Processing}

\subsubsection{Dataset} We utilize the HumanEval-X benchmark~\cite{zheng2023codegeex}, focusing on the JavaScript and Python subsets due to their prevalence in code obfuscation. HumanEval-X is ideal for output prediction studies because its programs are typically short, self-contained, and execution-focused, allowing participants to mentally simulate logic within a controlled timeframe. The benchmark's competitive-programming style emphasizes precise data flow and algorithmic reasoning, the core skills output prediction intends to measure, while minimizing the need for the domain-specific knowledge found in larger projects. From each language subset, we randomly selected 10 snippets subject to two constraints: a cyclomatic complexity between $4$ and $8$, and a line count of fewer than $15$. These criteria ensure that the base snippets are non-trivial yet concise, which is essential given that obfuscation significantly increases code difficulty.

\subsubsection{Code Obfuscation} For obfuscation tiers L1-L3, we generated variants using ObfuXtreme~\cite{obfuXtreme25} for Python and \code{javascript-obfuscator}~\cite{javascript_obfuscator_411} for JavaScript, ensuring all snippets remained functionally equivalent through execution testing with identical input arguments. 
For this study, we configured ObfuXtreme so that ControlFlowFlattener and VariableRenamer can be enabled independently, and we disabled StringEncryptor. We also shortened renamed identifiers by reducing the SHAKE-128 output from 8 to 2 hex characters, yielding names of the form \textit{var\_xxxx} (<=8 characters); variables introduced by control-flow flattening were not subject to this limit. For JavaScript, we used \code{javascript-obfuscator} with only the transformations needed for our tiers: L1, L2, and L3.
All other options were disabled to avoid unrelated transformations. After obfuscation, we replaced the function name with a neutral placeholder (e.g., \code{myFunct}) except adversarial renaming.

For L1b (adversarial renaming), we harvested identifiers from the test split (to avoid data leakage) of CodeSearchNet~\cite{husain2019codesearchnet}, which is large, available in both languages, and consists of function-level code, matching the scope of HumanEval-X. Using tree-sitter, we built four identifier pools: {JavaScript/Python} $\times$ {function names/variable names}. We replaced each function and variable name with randomly sampled name from the respective pool. To ensure semantic plausibility in context, we classified names into domains and picked all the names for a code snippet in the same domain.

\subsection{Human Study Design}

\begin{wraptable}{r}{3.2in}
\centering
\small
\caption{Self-reported JavaScript and Python experience.}
\label{tab:experience_grid}
\vspace{-9pt}
\begin{tabular}{lcccc}
\toprule
\textbf{JavaScript} & \multicolumn{3}{c}{\textbf{Python Experience}} & \textbf{Total} \\
\cmidrule(lr){2-4}
 {\bf Experience} & None & $<1$ year & $\geq$1 year & \\
\midrule
None         &  5 & 10 &  1 & 16 \\
$<1$ year    &  8 &  7 &  8 & 23 \\
$\geq$1 year &  1 &  4 &  6 & 11 \\
\midrule
\textbf{Total} & 14 & 21 & 15 & \textbf{50} \\
\bottomrule
\end{tabular}
\end{wraptable}

We administered an in-person survey via Qualtrics, conducted in a single room under the supervision of three researchers. All participants began simultaneously and were allotted 75 minutes to complete 12 questions. We prohibited communication, collaboration, and the use of all external resources, including search engines, LLMs, code execution tools, and IDEs. Each participant received blank A4 paper and a pencil for scratch work.

We recruited 50 undergraduate computer science majors through a course subject pool. We recorded each subject's self-reported Python and JavaScript experience and their per-question completion time. Experience was reported in three categories for each language: None, <= 1 year, and >1 year (Table~\ref{tab:experience_grid} shows the distribution across joint-experience groups). Completion time was measured in seconds via Qualtrics page-timer ``page submit'' metrics.

The survey consisted of a mixture of the Python or JavaScript questions. A participant answered questions randomly drawn from a pool of code snippets. For each snippet, participants saw exactly one of five obfuscation tiers (L0, L1, L1b, L2, L3). We balanced the assignment so that questions and tiers appeared evenly across participants. No participant saw the same base snippet in multiple obfuscation tiers, ensuring that obfuscation effects were isolated and preventing memory effect due to repeated exposure. Overall, each participant answered 12 output-prediction questions.

Each question followed the format: ``{\em Given the following [Python/JavaScript] function and input, what is the correct output}?'' The input was provided as a function call with specified argument(s) and was held constant across obfuscation variants of the same base question (Fig.~\ref{fig:obfuscation-levels}). A response was marked correct if it matched the output produced by executing the snippet with the given function call and arguments; near-matches were additionally adjudicated by human judgment.

\section{Impact of obfuscation on output prediction performance (RQ1)}

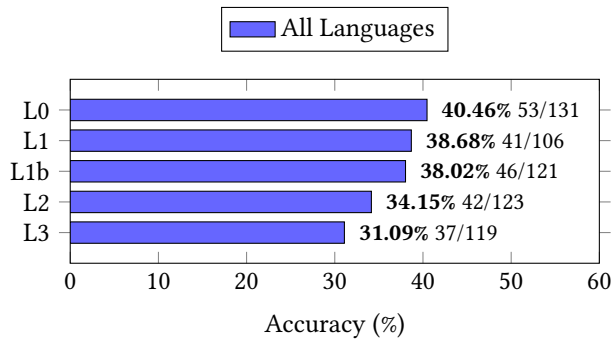
\begin{wrapfigure}{r}{0.62\textwidth}
\begin{tikzpicture}
\begin{axis}[
    xbar,
    xmin=0,
    xmax=60,
    width=\linewidth,
    height=4cm,
    bar width=8pt,
    enlarge y limits=0.25,
    xlabel={Accuracy (\%)},
    symbolic y coords={
        L0,
        L1,
        L1b,
        L2,
        L3
    },
    ytick=data,
    y dir=reverse,
    nodes near coords,
    nodes near coords align={horizontal},
    point meta=explicit symbolic,
    every node near coord/.append style={
        anchor=west,
        font=\small,
        xshift=2pt
    },
    legend style={
        at={(0.5,1.15)},
        anchor=south,
        legend columns=-1,
        draw=black,
        fill=white
    },
    legend image code/.code={
        \draw[#1, draw=black] (0cm,-0.1cm) rectangle (0.6cm,0.1cm);
    }
]

\addplot[
    fill=blue!60,
    draw=black
] coordinates {
    (40.46,L0)  [{\textbf{40.46\%} 53/131}]
    (38.68,L1)  [{\textbf{38.68\%} 41/106}]
    (38.02,L1b) [{\textbf{38.02\%} 46/121}]
    (34.15,L2)  [{\textbf{34.15\%} 42/123}]
    (31.09,L3)  [{\textbf{31.09\%} 37/119}]
};

\addlegendentry{All Languages}

\end{axis}
\end{tikzpicture}
\vspace{-12pt}
  \caption{Impacts of Different Obfuscation Tiers on Output Prediction. Bar labels report accuracy and (Correct / Total Attempts).}
  \label{fig:rq1}
\end{wrapfigure}

Fig.~\ref{fig:rq1} shows that the highest accuracy occurs under {\em L0} (40.46\%). With unobfuscated code, readers can leverage System 1 processing by relying on pattern matching of familiar constructs, loop shapes, and data-flow idioms. This intuitive scaffolding minimizes the need for users to ``slow down'' for full semantic parsing. However, while a descriptive name like \code{bubblesort} would allow users to bypass code reading entirely, all functions in our dataset (except L1b) use the generic identifier \code{myFunct}. Consequently, even the unobfuscated baseline demands a baseline, of System 2 reasoning.

In contrast, {\em Renaming + Control-Flow} had the lowest accuracy (31.09\%), confirming combined obfuscations compound difficulty. By simultaneously removing semantic shortcuts (System 1) and forcing laborious mental simulation with control-flow transportation (System 2), this combination maximizes cognitive load, leading to fatigue, mistakes, and tracing errors.

The results indicate that {\em humans are more resilient to renaming than to control-flow obfuscation}, with the largest accuracy drop ($\Delta -3.87\%$) occurring between L1b (Adversarial Renaming) and L2 (Control-Flow). While adversarial renaming weakens semantic cues, it preserves the underlying execution structure, allowing readers to reconstruct meaning through partial heuristics (e.g., ``this variable behaves like an accumulator'').  Control-flow obfuscation eliminates these structural regularities (recognizable branches/loops), thereby removing System 1 crutches and forcing a switch to laborious System 2 tracing.

Analysis of user responses to the Tribonacci task (Fig.~\ref{fig:obfuscation-levels}) supports this. Calculating \code{myFunct}(14) requires recursively summing terms to reach 927, a task exceeding working memory. If System 2 was engaged, answers should be close to 927. However, out of twelve L0 and L1b responses, only one was correct. The majority of incorrect responses were heuristic guesses anchored to the input (12, 15) or small Fibonacci numbers (0, 1, 3, 5), indicating a failure to transition from intuitive System 1 to analytical System 2. In contrast, when obfuscation alters literals (L2/L3), the errors shift from numeric guesses to syntax confusion. Readers struggled with hexadecimal numbers, frequently outputting strings like "0x0" or "0x1," or simply stating "I don't know." Overall, L0 and~L1b errors are rectifiable by a transition to System 2, whereas L3 errors reflect a collapse in the ability. 

Finally, the overall performance level (36.5\% correct out of 600 total responses) shows that the tasks were challenging even without obfuscation (baseline is only 40.46\%). Thus, obfuscation is reducing accuracy from an already nontrivial comprehension setting, and the decrease to ~31.09\% (L3) represents a meaningful loss in successful understanding at scale. In Dual-System Theory terms, that suggests many questions already required meaningful System 2 engagement such as careful tracing. Obfuscation then acts like an added ``tax'' that (1) reduces the availability of System 1 shortcuts and (2) increases the amount of System 2 work required. The net effect is a noticeable drop in correctness--especially when control flow is distorted--because more people rely on System~2.

\begin{tcolorbox}[colback=white, colframe=black, arc=8pt, boxrule=0.5pt]
    \textbf{RQ1 Takeaway:} \textit{We observe decreased accuracy across all obfuscation tiers. Variable Renaming and Adversarial Renaming show a smaller decrease in accuracy compared to the baseline of -1.78\% and -2.44\% respectively. Control Flow and Renaming+Control Flow had much stronger effects on accuracy overall, -6.31\% and -9.37\% respectively.
}
\end{tcolorbox}

\section{How does increasing code obfuscation levels ($L0$--$L3$) affect "time to complete" the
output prediction tasks by humans? (RQ2)}

\begin{figure}[t]
  \centering
  \begin{minipage}[t]{0.42\textwidth}
    \centering
    \includegraphics[width=0.9\linewidth]{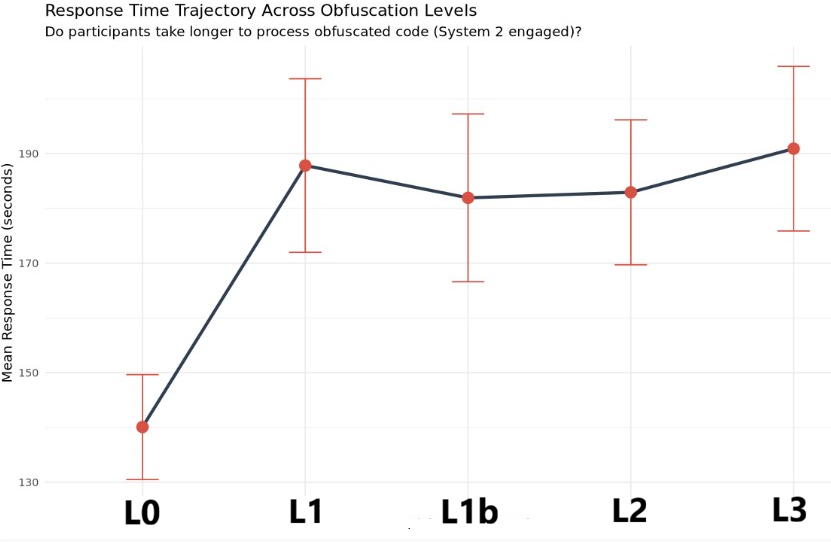}
    \vspace{-7pt}
    \caption{Response Time Across Obfuscation}
    \label{fig:response-time}
  \end{minipage}\hfill
  \begin{minipage}[t]{0.42\textwidth}
    \centering
    \includegraphics[width=0.9\linewidth]{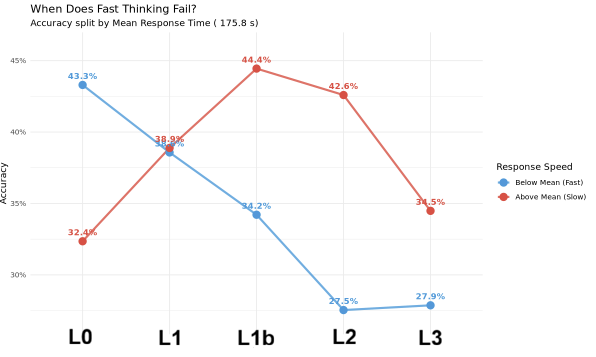}
    \vspace{-7pt}
    \caption{When does Fast-Thinking Start Failing?}
    \label{fig:mean-division}
  \end{minipage}
\end{figure}

We map response times to the Dual-Process model: shorter latencies represent System 1 (heuristic/pattern recognition), while longer latencies represent System 2 (deliberate tracing and verification).

\textbf {Time to response distribution}. Figures~\ref{fig:response-time} and~\ref{fig:mean-division} demonstrate a clear shift in strategy when obfuscation is introduced. For original code ($L0$), participants rely on System 1 fluency, responding fastest (avg. 140s) with high accuracy. Once obfuscated ($L1$--$L3$), mean response times jump by approximately one minute and remain elevated. This indicates that participants must abandon rapid pattern recognition for sustained, effortful System 2 tracing. As shown in Fig.~\ref{fig:mean-division}, the accuracy of "fast" responses (blue) declines steadily as obfuscation levels increase, suggesting that System 1 heuristics no longer match the code's surface cues. In contrast, the "slow" group (red) maintains higher accuracy. $L2$ shows the largest accuracy gap ($\Delta$15.1\%), indicating it consistently confuses System 1 readers but remains tractable for those who invest time in System 2 tracing. The most striking case is L1b, where slow responses peak in accuracy, suggesting a ``sweet spot'' where obfuscation is challenging enough to trigger careful attention but still tractable enough that extra effort pays off. L1b and L2 are the tasks that most benefit from System 2 thinking. L1b has the highest overall accuracy for the red graph, showing that once users activate System 2 thinking, it is most comprehensible to users. This squares with the hypothesis that humans are able to understand code with adversarial renaming as long as they are not initially deceived by the semantically misleading names. L2 shows the largest red-blue accuracy gap (\(\Delta \)15.1\%), indicating it confuses System~1 readers but is better understood with more time.
For the L2 level, the largest accuracy gap between the accuracy of the red and blue graphs (\(\Delta \)15.1\%), indicating that this level confused participants using System 1 Thinking but was better understood by participants with longer response times.
At L3, however, accuracy drops even for the slow group (34.5\%, near the baseline 32.4\%), suggesting that renaming+control-flow obfuscation overwhelms comprehension despite System~2 effort.

\begin{wrapfigure}{r}{0.48\textwidth}    \centering
    \includegraphics[width=0.48\textwidth]{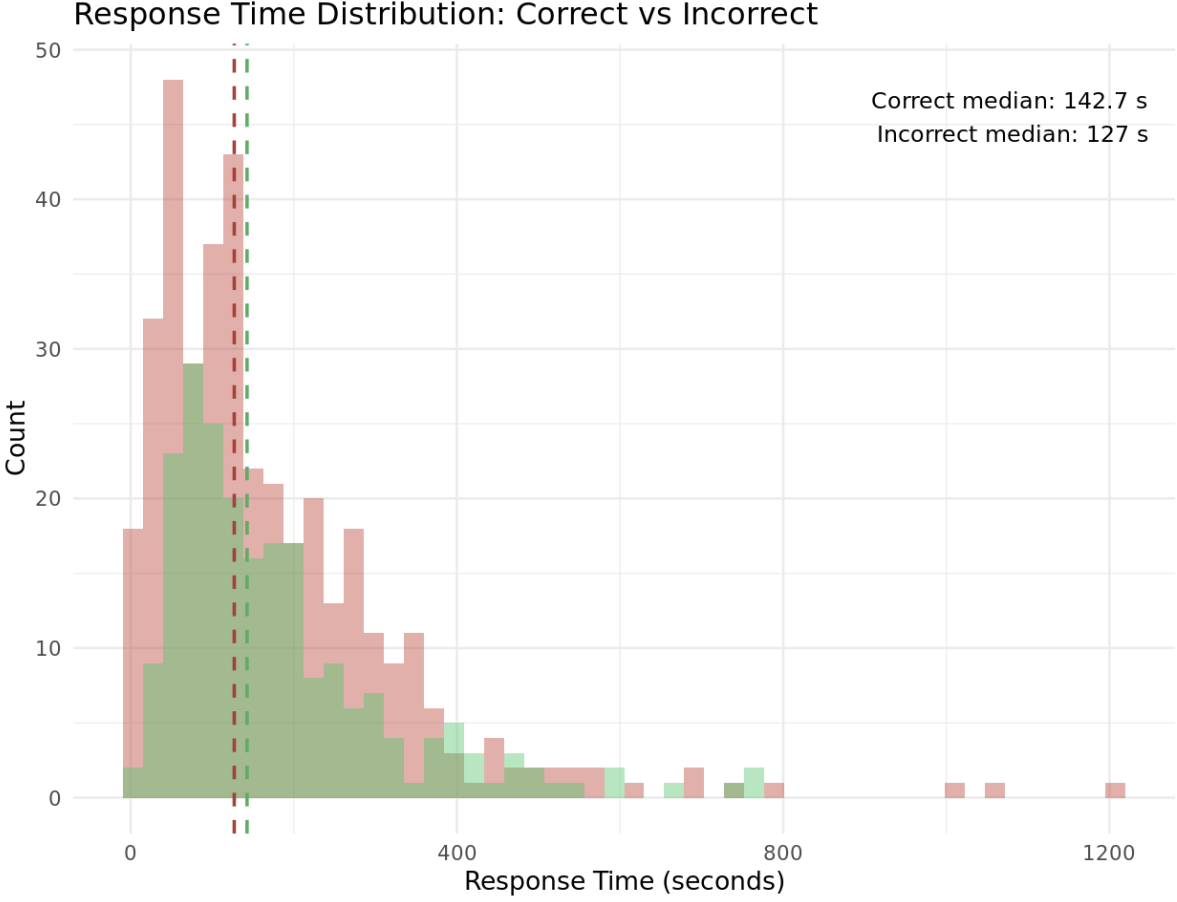}
    \vspace{-15pt}
    \caption{Response time distribution of correct (Green) responses and incorrect responses (Red).}
    \label{fig:correct-vs-incorrect-times}
\end{wrapfigure}

\textbf {Stratifying time-to-responses by correct and incorrect answers}. Fig.~\ref{fig:correct-vs-incorrect-times} shows {\bf right-skewed} response-time distributions for both correct and incorrect answers, indicating substantial variability in how long participants needed to process the code--most responses cluster at shorter times, with a long tail of much slower decisions. This pattern aligns with Dual-System theory: shorter latencies often reflect System 1 (fast, heuristic) processing, while longer latencies suggest greater System 2 (slow, deliberate) engagement. The incorrect distribution is more focused in the 0--50s range, implying many errors stem from rapid heuristic judgments that break under complex obfuscation.

In contrast, correct responses rise sharply around 50s, suggesting a cognitive floor for System 2 tracing; responses below this threshold are likely too fast to have correctly followed the program.
Moreover, both distributions have a long tail: some participants spent several minutes (up to 20 of 75 minutes) on a single response, consistent with System 2 tracing and verification. However, the most extreme times occur mostly among incorrect answers, suggesting that beyond a point extra time reflects confusion or unproductive search (e.g., repeated re-checking or getting lost in control flow), not improved accuracy--{\em System 2 engagement is not automatically beneficial}. 

The incorrect outliers in Fig.~\ref{fig:correct-vs-incorrect-times} are three attempts (three participants), all on Python questions at different tiers (L1, L1b, L3) that include identifier renaming; their self-reported Python experience is ($\ge$1 year, $<1$ year, $<1$ year), respectively, and all are incorrect, suggesting that extreme response time may be associated with difficulty in moving beyond surface-level cues.

Two other observations reinforce this interpretation. First, the medians (142.7s correct vs. 127s incorrect) show that {\em correct answers tend to be somewhat slower}, consistent with successful performance requiring System 2 checking--especially since most questions are obfuscated and demand time to move beyond surface-level cues. Second, the overlap between histograms suggests no single ``optimal'' time; instead, performance follows a Goldilocks pattern: very fast responses are more error-prone (System 1), moderate deliberation improves accuracy (effective System 2; correct answers cluster around 60--180s for JavaScript and 100--300s for Python), and extremely long times increasingly reflect breakdowns in comprehension (System 2 struggling without converging).

\begin{wrapfigure}{r}{3.3in}
    \centering
    \includegraphics[width=0.6\textwidth]{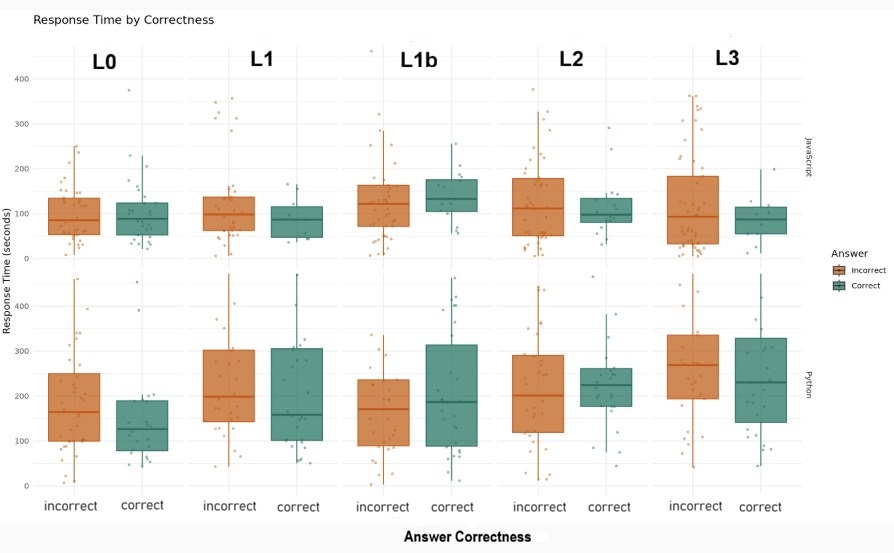}
    \caption{Response time distributions for correct (green) and incorrect (orange) responses by obfuscation tier.}
    \label{fig:response_time_by_correctness}
\end{wrapfigure}

\textbf{Stratification on different obfuscation levels} As seen~in Fig.~\ref{fig:response_time_by_correctness}, across most obfuscation levels, {\em correct responses tend to be faster than incorrect ones}: the ``Correct'' group typically has a lower median response time (often with a tighter spread), while incorrect answers show wider dispersion and longer upper whiskers/outliers. A plausible interpretation is that more fluent programmers can quickly form an accurate mental model and respond confidently, whereas confusion drives time upward--participants trace, backtrack, and re-check assumptions but fail to converge. 

This {\em slow-but-wrong} pattern is evident in~the heavier tails for incorrect responses, suggesting extra time often reflects difficulty than~correct~verification.
The notable exception is L1b (Adversarial Renaming), where correct responses are consistently slower: the ``Correct'' distributions sit above ``Incorrect'' in both language panels. This suggests adversarial renaming especially punishes fast, heuristic processing--quick responders may trust misleading surface cues (e.g., corrupted variable-name semantics) and answer confidently but incorrectly. Those who slow down instead trace dataflow and variable usage rather than names, which takes longer but yields more correct answers. A few additional patterns stand out.

First, as obfuscation increases (especially toward L2/L3), variance grows and outliers become more extreme, suggesting a split between participants who can still cope efficiently and those who get stuck. Second, the Python panel shows generally higher response times and wider spreads than JavaScript, implying greater difficulty or stronger sensitivity to obfuscation.

Response time is context-dependent rather than uniform: in L0 and L1, speed mostly signals fluency; under adversarial renaming, slower responses reflect successfully overriding deceptive intuitions; and in L2/L3, very fast responses are often wrong while the longer times indicate confusion, with correct answers clustering in a middle band that reflects deliberate yet efficient~reasoning.

\textbf{The Non-Monotonic Relationship of Time and Accuracy.} Tiers are classified by their dominant cognitive mode: System 1-aligned if the fastest responses are most accurate, or System 2-dependent if accuracy scales with deliberation time.

\begin{figure}[t]
\centering
\small
\begin{minipage}[t]{0.48\textwidth}
    \centering
    \includegraphics[width=\linewidth]{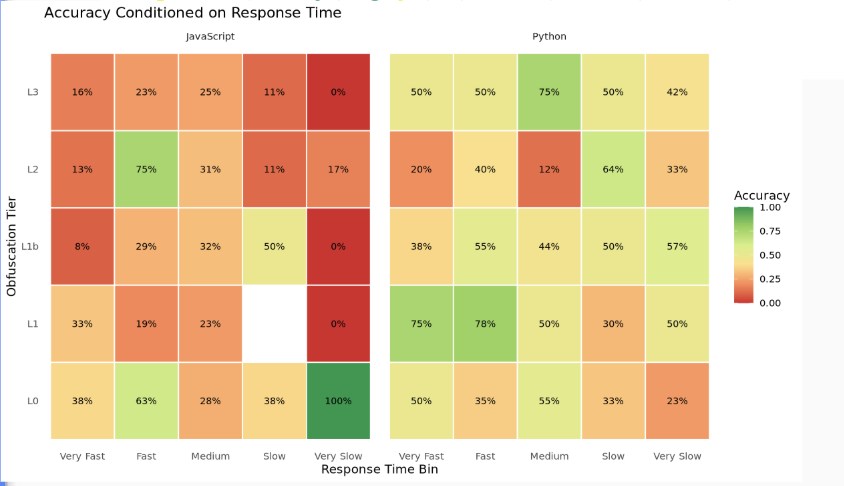}
    \vspace{-12pt}
    \captionof{figure}{Heatmap of Accuracy conditioned on Response Time and Obfuscation Tier. The visualization displays mean accuracy rates across five response time bins (Very Fast to Very Slow) and five obfuscation tiers ($L0$--$L3$) for JavaScript and Python.}
    \label{fig:heatmap}
\end{minipage}\hfill
\begin{minipage}[t]{0.48\textwidth}
    \centering
    \includegraphics[width=\linewidth]{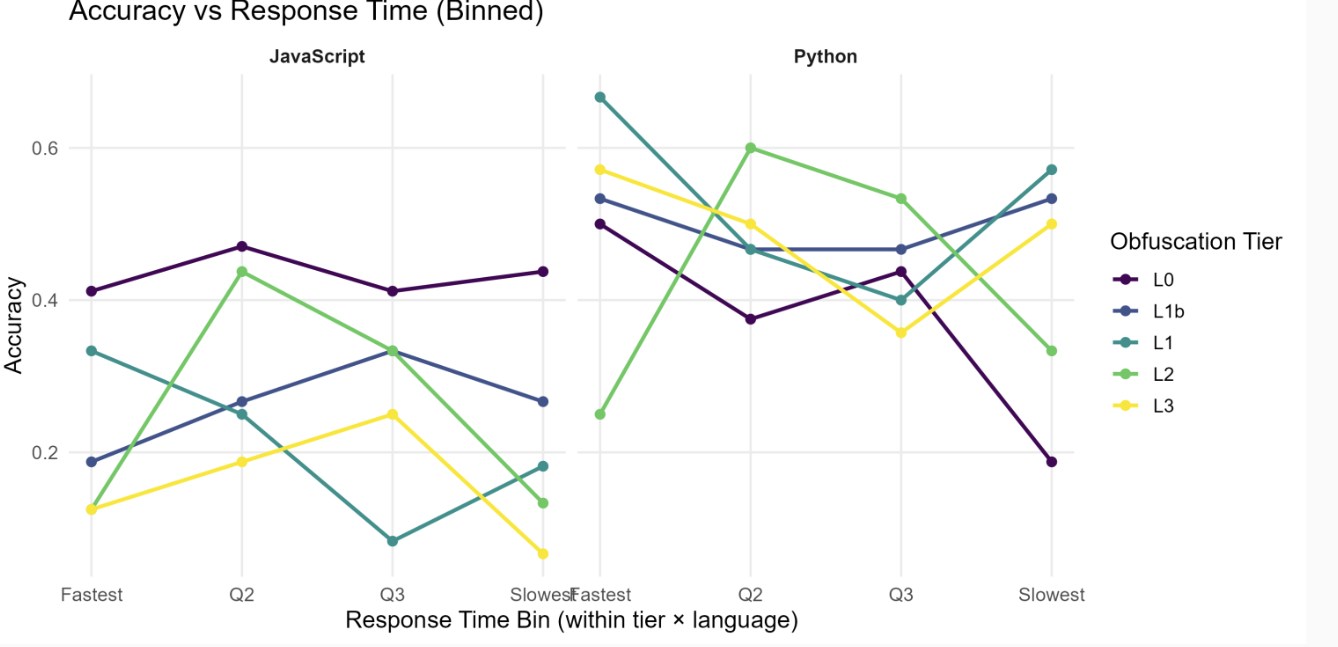}
    \vspace{-12pt}
    \captionof{figure}{comparison of participant accuracy against response speed bins for each obfuscation tier. The data is split between JavaScript (left) and Python (right) to highlight language-specific performance variances across time quartiles.}
    \label{fig:linegraph-time-obfuscation}
\end{minipage}
\end{figure}

For {\bf System 1-aligned cases (Python L0 \& L1, JavaScript L1)}, the pattern is essentially "you know it or you don't." As seen in both the Heatmap (Fig.~\ref{fig:heatmap}) and Linegraph (Fig.\ref{fig:linegraph-time-obfuscation}), accuracy is highest in the fastest bins and drops steadily as responses slow down (e.g., Python L1 drops from 78\% to 30\%). This suggests that for unobfuscated or lightly obfuscated code, correct answers stem from rapid recognition; prolonged time here is a marker of uncertainty or lack of fluency rather than effective reasoning.

For {\bf System 2-aligned cases (all other obfuscation tiers)}, the trend flips: fast responses are generally poor and accuracy tends to improve with more time, consistent with the need for System 2--style tracing. In {\bf Python L2}, accuracy rises from 20\% (very fast) and 40\% (fast) to a peak at slow (64\%), then drops in the very slow bin (33\%), suggesting unproductive struggle. In {\bf JavaScript L3}, accuracy remains low but still improves from very fast (16\%) to medium (25\%), before collapsing at very slow, again indicating that extreme time often reflects confusion. 
Moreover, {\bf Python appears more resilient overall}: for instance, Python L0 shows high accuracy even for very fast/fast responses (75--78\%), while {\bf JavaScript L1} is much lower (33\% and 19\%). L1b also shows a stronger ``slowing helps'' signature in JavaScript than in Python: JavaScript L1b peaks in the slow bin (50\%) but falls to near zero in the very slow bin, suggesting adversarial semantics triggers deliberate checking, while extreme deliberation can still fail if the mental model never~stabilizes.

From Fig.~\ref{fig:linegraph-time-obfuscation}, several tiers show non-monotonic U-shaped or inverted-U patterns rather than ``slower is better.'' For instance, L1 drops sharply from the fastest bin to Q3 and only slightly rebounds at the slowest bin, suggesting extra time often reflects getting stuck. L3 improves from fastest to Q3 but collapses at the slowest bin, indicating very slow responses can signal confusion even when moderate deliberation helps. This plot also clarifies the ``Goldilocks zone'': for harder tiers in both languages, accuracy often peaks in the middle bins (Q2/Q3), while the slowest bin often dips (e.g., JavaScript L3; Python L2), consistent with moderate System~2 engagement helping but excessive time reflecting breakdown. Compared to JavaScript, Python curves are generally higher and smoother (especially L1/L1b/L3), whereas JavaScript shows sharper collapses (e.g., L1 at Q3, L3 at slowest), suggesting obfuscation more often derails readers into unproductive traces.

\begin{tcolorbox}[colback=white, colframe=black, arc=8pt, boxrule=0.5pt]
    \textbf{RQ2 Takeaway:} \textit{Obfuscation disrupts intuitive guessing (System 1), forcing users to manually verify logic flow (System 2), while rapid thinking succeeds on clean code ($L0$) since the participants can rely on code heuristics and signposts, this cognitive style leads to errors in obfuscated tiers. Adversarial Renaming shows the highest accuracy under System 2. Time~shares a non-linear relationship with accuracy: results are poor at both extremes (hasty errors vs. prolonged confusion), with peak performance occurring only during moderate intervals of System 2 engagement. Fast can work when tasks are easy, moderate slowing helps when tasks are hard, and extremely slow responses often indicate breakdown rather than successful reasoning.
}
\end{tcolorbox}
\section{Impact of code complexity on performance (RQ3)}

\textbf{ Identifier Length vs. Accuracy.} We found no significant correlation between identifier length and performance, challenging the assumption that verbosity implies clarity. Although short variables (e.g., \textit{x}) are often less interpretable than descriptive ones (e.g., \textit{calculateTotal}), adversarial renaming breaks this relationship: misleading identifiers can be long yet still harm comprehension. This added noise scatters the results, suggesting identifier veracity matters more than identifier length.

\begin{wrapfigure}{r}{0.45\textwidth}
    \centering
    \includegraphics[width=0.38\textwidth]{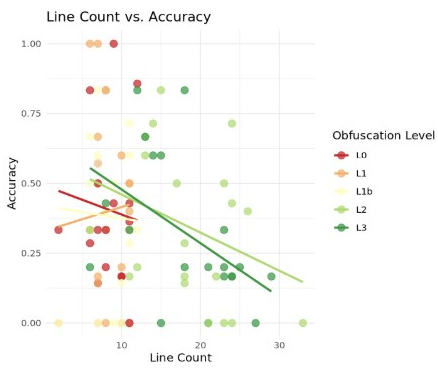}
    \vspace{-12pt}
    \caption{Line Count vs. Accuracy by Obfuscation}
    \label{fig:code_length_accuracy}
\end{wrapfigure}

\textbf{ Line Count vs. Accuracy.} As shown in Fig.~\ref{fig:code_length_accuracy}, accuracy generally declines as line count increases. Variability is high for snippets under 10 lines, suggesting item difficulty dominates at small sizes (some are obvious, others deceptively tricky). L2 and L3 drop most steeply, indicating control-flow transformations scale cognitive cost with program size: longer code adds more blocks and dependencies, forcing sustained System~2 tracing and making structural disruption more damaging.

L1 is the notable exception: accuracy increases with line count. Under identifier renaming, name-based System~1 cues are degraded, and very short snippets offer too little structure to recover intent, pushing readers into System~2 without enough context. Longer code provides more structural cues (e.g., loop boundaries, nesting), helping readers reconstruct meaning despite uninformative names. For example, in Python No.101, L0 responses captured the string content (``Hi my name is John'') but missed formatting it as a list; in L1, renaming variables (e.g., \code{letter} to \code{var\_bc5a}) forced users to verify syntax rather than assume the output, improving accuracy.

\begin{figure}[!t]
\centering
\small
\begin{minipage}[t]{0.48\textwidth}
    \centering
    \includegraphics[width=\linewidth]{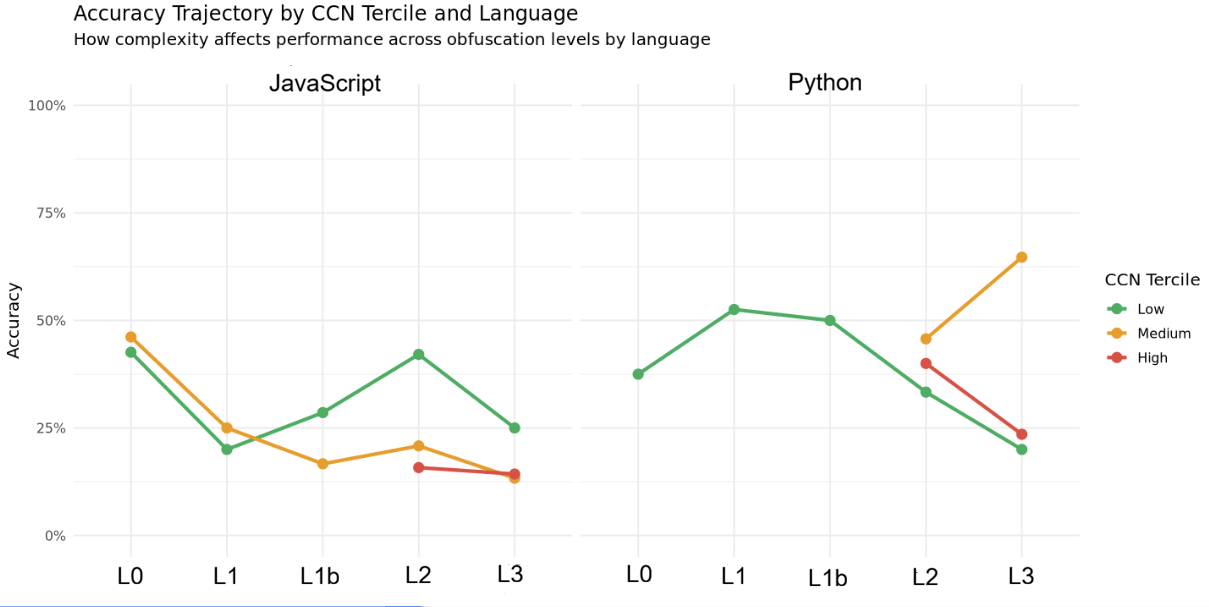}
    \vspace{-15pt}
    \captionof{figure}{Accuracy vs. Cyclo. Complexity by Obfus.}
    \label{fig:acc_cc_tercile_language}
\end{minipage}\hfill
\begin{minipage}[t]{0.48\textwidth}
    \centering
    \includegraphics[width=\linewidth]{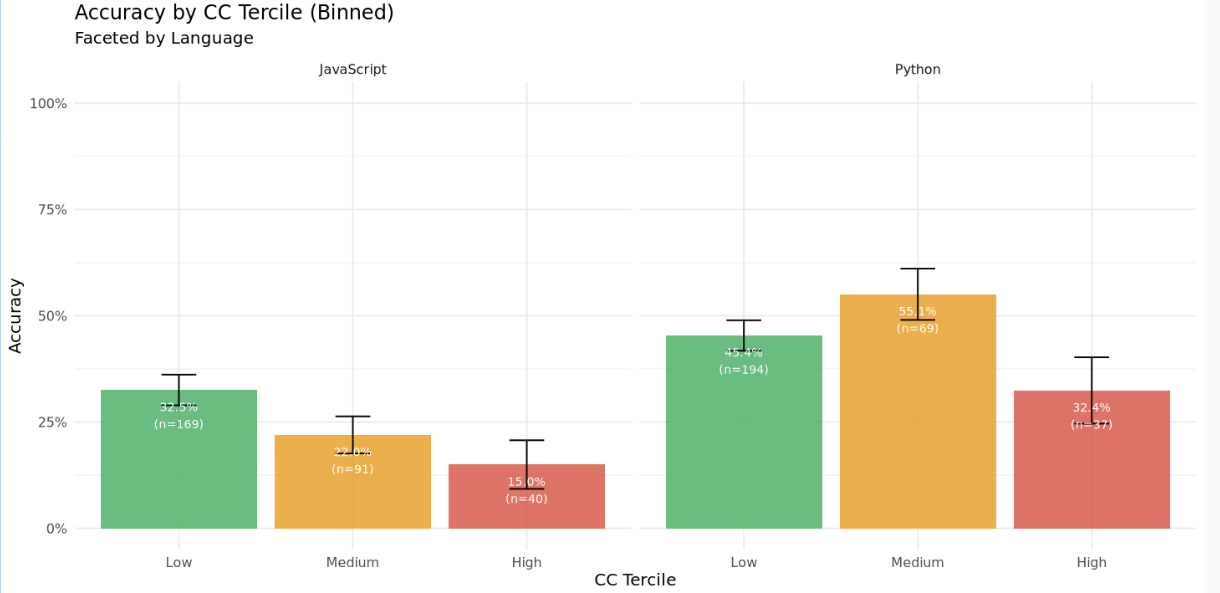}
    \vspace{-15pt}
    \captionof{figure}{Accuracy vs Cyclo. Complexity by Language}
    \label{fig:acc_by_cc_language}

\end{minipage}
\end{figure}

\textbf{Accuracy Trajectory by Cyclomatic Complexity (CC) for Different Obfuscation Levels.} 
Following McCabe, we use Cyclomatic Complexity as a proxy~\cite{mccabe_cc}. Fig.~\ref{fig:acc_cc_tercile_language} shows that obfuscation interacts with the baseline complexity (line count and code complexity are measured on original code to avoid conflation). For High CC (red), accuracy declines monotonically in both languages (L3 < L2 < L1), indicating that when code is already complex, obfuscation simply compounds difficulty without beneficial strategy shifts; the drop is steeper in Python (especially L2 to L3).

In contrast, Low and Medium CCN show non-monotonic ``dip-rebound'' patterns. Most notably, Python Medium CCN improves from L2 to L3 (mid-40\% to mid-60\%), suggesting that adding renaming atop control-flow alteration can sometimes aid comprehension--perhaps by discouraging misleading semantic cues and prompting more systematic tracing. Another possibility is that the specific subset of medium-CCN items reaching L3 may have more consistent structure or be more ``traceable'' despite obfuscation--either way, L3 is not simply ``harder'' here. Low CCN shows a similar ``beneficial difficulty'' effect (Python peaks at L1/L1b; JavaScript at L2), but this disappears at High CCN, where complexity plus obfuscation overwhelms tracing even with increased System~2 effort.

\textbf{ Accuracy by Cyclomatic Complexity.} As shown in Fig.~\ref{fig:acc_by_cc_language}, JavaScript accuracy decreases monotonically with cyclomatic complexity, most sharply in the High CC bin, as more paths and decision points increase mental execution burden and errors. Python is non-monotonic: Medium complexity appears to provide structural scaffolding (e.g., intermediate variables, staged computations) that supports System~2 tracing, unlike Low CC where compact idioms can trigger misleading heuristics and High CC which overwhelms working memory. Participants perform better on Python than JavaScript at every CC tier, suggesting either the Python items are more traceable in this dataset or participants have higher baseline Python fluency.

\begin{tcolorbox}[colback=white, colframe=black, arc=8pt, boxrule=0.5pt]
    \textbf{RQ3 Takeaway:} \textit{
    Complexity affects accuracy non-monotonically. Misleading names disrupt comprehension regardless of their length. Control-flow obfuscation becomes increasingly harmful as programs get longer, whereas pure renaming can sometimes be offset by longer structure that provides more traceable cues. Cyclomatic complexity is a strong predictor of difficulty in JavaScript, while in Python the relationship is more nuanced, with medium-complexity code sometimes being the most traceable for humans, suggesting that readers sometimes require more logic, not less, to anchor their reasoning when surface cues are removed.  
}
\end{tcolorbox}

\section{Impact of programming language on performance (RQ4)}

\begin{wrapfigure}{r}{0.60\textwidth}
\centering
\begin{tikzpicture}
\begin{axis}[
    xbar,
    xmin=0,
    xmax=55, 
    width=\linewidth,
    height=4cm,
    bar width=5pt,
    enlarge y limits=0.3,
    symbolic y coords={
        None,
        $<1$ year,
        $\geq$1 year
    },
    ytick=data,
    y dir=reverse,
    nodes near coords,
    nodes near coords align={horizontal},
    point meta=explicit symbolic,
    every node near coord/.append style={
        anchor=west,
        font=\scriptsize,
        xshift=2pt
    },
    legend style={
        at={(0.5,1.15)},
        anchor=south,
        legend columns=-1,
        draw=black,
        fill=white,
        /tikz/every even column/.append style={column sep=0.5cm}
    },
    legend image code/.code={
        \draw[#1, draw=black] (0cm,-0.1cm) rectangle (0.6cm,0.1cm);
    }
]

\addplot[
    fill=blue!60,
    draw=black
] coordinates {
    (30.95,None) [{\textbf{30.95\%} 52/168}]
    (38.89,$<1$ year) [{\textbf{38.89\%} 98/252}]
    (38.33,$\geq$1 year) [{\textbf{38.33\%} 69/180}]
};
\addlegendentry{Python}

\addplot[
    fill=orange,
    draw=black
] coordinates {
    (35.94,None) [{\textbf{35.94\%} 69/192}]
    (36.23,$<1$ year) [{\textbf{36.23\%} 100/276}]
    (37.88,$\geq$1 year) [{\textbf{37.88\%} 50/132}]
};
\addlegendentry{JavaScript}

\end{axis}
\end{tikzpicture}
\vspace{-21pt}
\caption{Output prediction accuracy by experience and language.}
\label{fig:acc_by_exp_lang}
\end{wrapfigure}
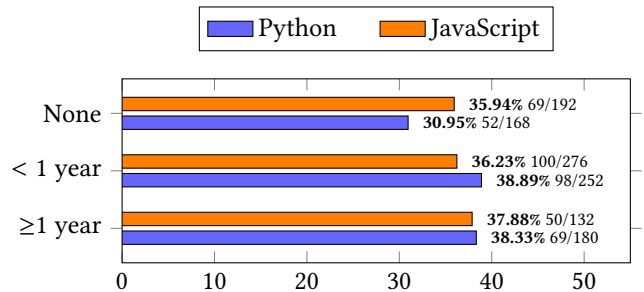

\textbf{Experience and Accuracy}. Fig.~\ref{fig:acc_by_exp_lang} shows the output prediction accuracy for both languages for each experience level. For both languages, experience is relatively spread out, so results are never driven by a tiny subgroup. For Python, the main gap is between participants with no Python experience and those with any experience, consistent with modest familiarity providing System~1 scaffolding (syntax fluency and construct recognition) and reducing System~2 effort spent decoding the language.
Interestingly, participants with > 1 year of experience do marginally worse than those with <=1 year (38.33\% vs. 38.89\%). One plausible explanation is a ceiling effect: beyond minimal fluency, additional experience yields diminishing returns, and performance becomes constrained by other factors (e.g., careful tracing demands, working-memory limits, or trickier semantic reasoning) rather than Python-specific knowledge. Another possibility is that more experienced participants may rely more heavily on quick pattern-based expectations (System 1 ``I've seen this before''), which can occasionally misfire if the snippets are atypical, whereas less experienced participants may slow down and engage System 2 consistently. 

The data indicates that time investment is a more critical factor for accuracy than prior experience. On Python/40 L1b a user despite having over a year of experience answered incorrectly after spending only 60 seconds on a task where the population average was 160 seconds. In contrast, participant No.6 demonstrated that effort can supersede experience; despite having no background in JavaScript, they outperformed the population average by securing a 33\% accuracy rate. Their success was directly correlated with time spent, as they failed questions addressed in under a minute but succeeded on those where they invested over two minutes. These cases suggest that expertise is easily undermined by haste, while focused diligence allows even beginners to surpass typical performance benchmarks.

JavaScript shows a weaker experience--performance link than Python. Participants with no experience (35.94\%) performed nearly the same as those with <= 1 year (36.23\%) and only slightly below those with 
>1 year (37.88\%). This suggests success depends less on language-specific knowledge than on transferable code-reading skills: novices can apply general System~1 pattern recognition and System~2 tracing once basic parsing is possible. The small experience gain also hints at a ceiling effect, where correctness is driven more by deep reasoning and verification than syntax fluency.

\begin{figure}[h]
  \centering
  \begin{minipage}{0.48\textwidth}
\tabcolsep 2.2pt
\footnotesize
\begin{tabular}{llrrr}
\hline
Language & Tier & \# attempts & Correct atts & Acc.  \\
\hline
JavaScript & L0  & 67 & 29 & 0.433 \\
JavaScript & L1  & 47 & 10 & 0.213 \\
JavaScript & L1b & 61 & 16 & 0.262 \\
JavaScript & L2  & 62 & 16 & 0.258 \\
JavaScript & L3  & 63 & 10 & 0.159 \\
\hline
Python & L0  & 64 & 24 & 0.375 \\
Python & L1  & 59 & 31 & 0.525 \\
Python & L1b & 60 & 30 & 0.500 \\
Python & L2  & 61 & 26 & 0.426 \\
Python & L3  & 56 & 27 & 0.482 \\
\hline
\end{tabular}
\end{minipage}
  \begin{minipage}{0.48\textwidth}
    \centering
    \includegraphics[width=0.95\textwidth]{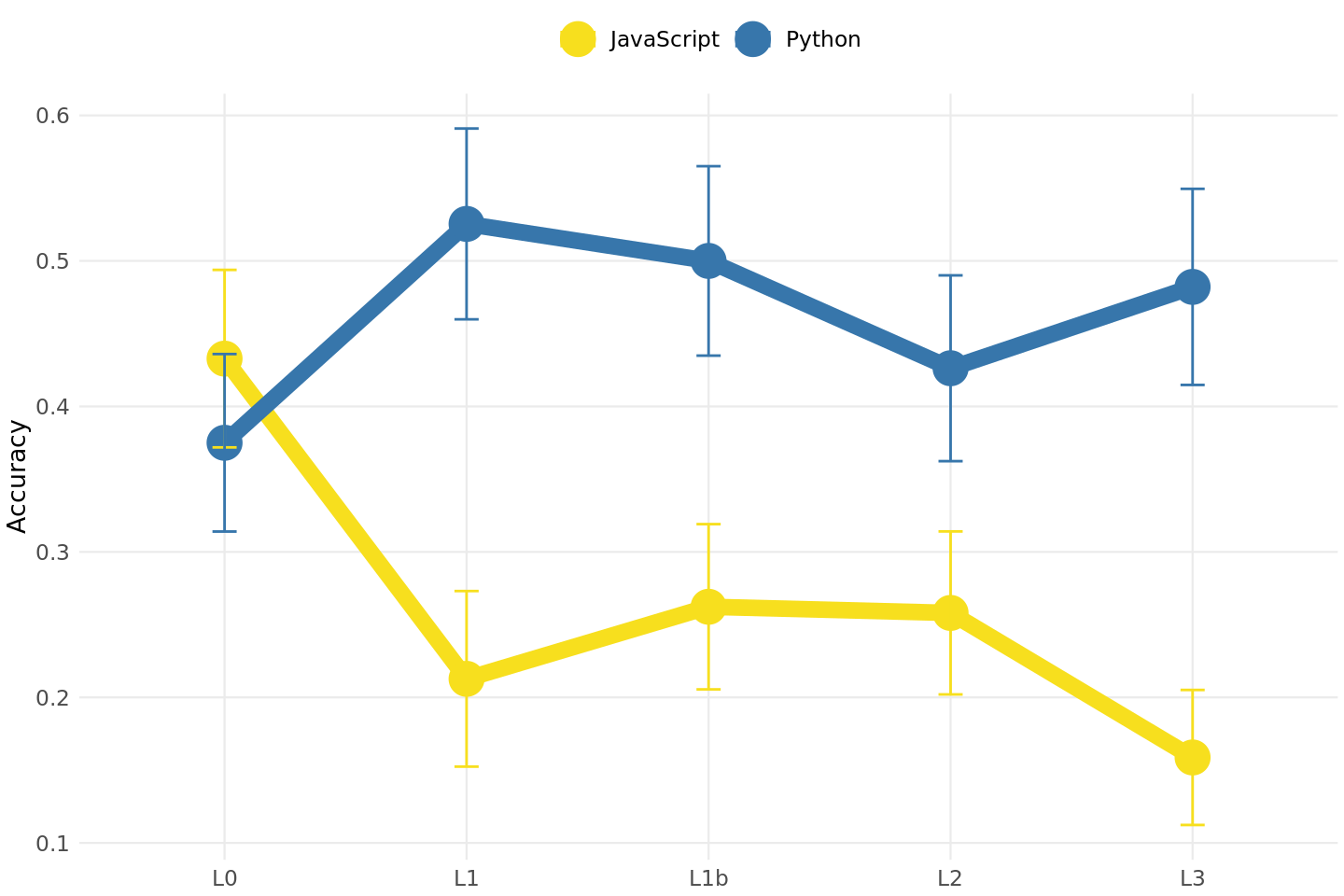}
\end{minipage}
\vspace{-9pt}
\caption{Accuracy by language and obfuscation, with tabular results (left) and accuracy visualization (right)}
\label{fig:accuracy_language_tier}  
\end{figure}

\textbf{Accuracy by Language and Obfuscation Tier.} As shown in Fig.~\ref{fig:accuracy_language_tier}, JavaScript accuracy declines with obfuscation intensity, dropping from 0.433 (L0) to 0.159 (L3), consistent with obfuscation eroding System~1 cues and forcing more error-prone System~2 tracing. However, participants did slightly better on Control Flow (L2, 0.258) than on Variable Renaming (L1, 0.213), suggesting identifiers are key anchors for forming a mental model in JavaScript. Interestingly, accuracy was higher under Adversarial Renaming (L1b, 0.262) than plain renaming, plausibly because semantically incongruent names trigger suspicion and prompt earlier System~2 checking.

Python shows a paradox: accuracy is lowest for unobfuscated code (L0, 0.375) and \textit{improves}~under renaming, with L1 (0.525) and L1b (0.500) highest. This suggests the original Python identifiers may prompt misleading System~1, domain inferences; stripping names forces more structure-based System~2 tracing. The control-flow results support this: L3 (renaming+control flow, 0.482) outperforms L2 (control flow, 0.426), implying renaming can act as a corrective even as complexity~increases.

Cross-language, identifier semantics seem to play different roles: in JavaScript, renaming hurts more than control-flow changes (L1 < L2), indicating heavier reliance on names, whereas in Python those cues may be counterproductive and their removal improves accuracy. Attempts per cell are broadly comparable ($\approx$ 47--67), though more testing is needed to confirm which gaps are~reliable.

\begin{figure}[t]
   \centering
   \begin{minipage}{0.48\textwidth}
   \footnotesize
    \tabcolsep 3pt
    \begin{tabular}{llrrrr}
   \toprule
 \textbf{Question Lang.} &
 \textbf{Obf.} &
 \textbf{\# Attempt} &
 \textbf{Corr. Att.} &
 \textbf{Acc.(\%)} \\
 \midrule
 JavaScript & L0  & 67 & 29 & 43.28 \\
            & L1  & 47 & 10 & 21.28 \\
            & L1b & 61 & 16 & 26.23 \\
            & L2  & 62 & 16 & 25.81 \\
            & L3  & 63 & 10 & 15.87 \\
 \midrule
 Python     & L0  & 64 & 24 & 37.50 \\
            & L1  & 59 & 31 & 52.54 \\
            & L1b & 60 & 30 & 50.00 \\
            & L2  & 61 & 26 & 42.62 \\
            & L3  & 56 & 27 & 48.21 \\
 \bottomrule
 \end{tabular}
   \end{minipage}\hfill
   \begin{minipage}{0.48\textwidth}
     \centering
     \includegraphics[width=0.75\textwidth]{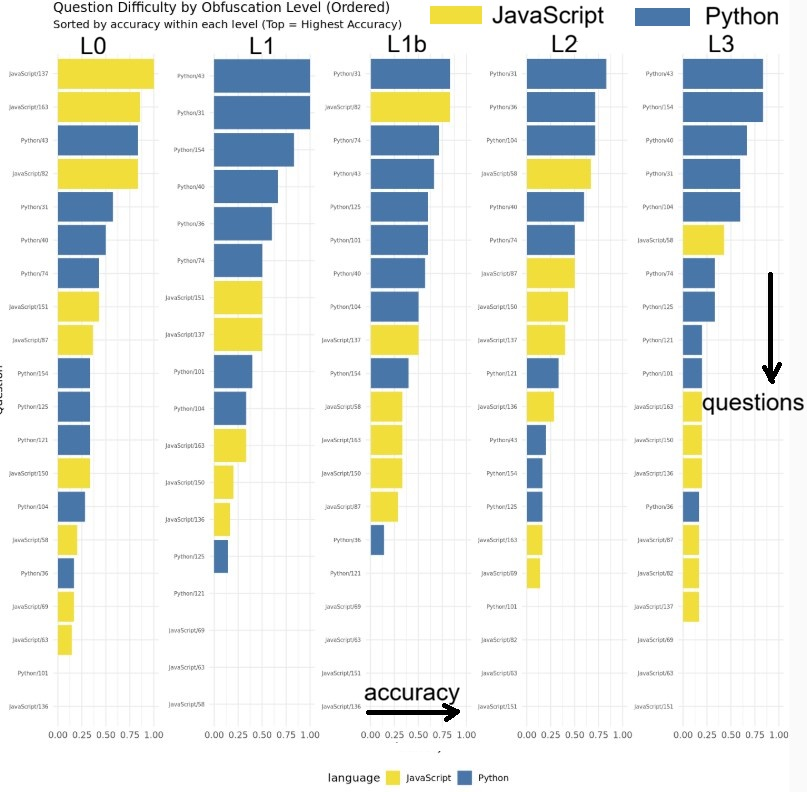}
 \end{minipage}
 \vspace{-9pt}
 \caption{Accuracy by question language and obfuscation tier}
 \label{fig:lang_tier_accuracy}  
\end{figure}

\textbf{Stratifying results on Questions and Obfuscation.} In Fig.~\ref{fig:lang_tier_accuracy}, Python yields higher average accuracy than JavaScript in every tier except L0. The unweighted averages show a clear gap (Python: 46.2\% vs. JavaScript: 26.5\%), mirrored in the weighted totals (46\% vs. 27\%); visually, Python sits above JavaScript across L1--L3, suggesting the easiest obfuscated items are popularly Python.

JavaScript declines monotonically with stronger obfuscation. Python is non-monotonic: accuracy rises from L0 (37.50\%) to L1 (52.54\%) and L1b (50.00\%), drops at L2 (42.62\%), then rebounds at L3 (48.21\%). This suggests renaming can reduce misleading semantic cues and push readers toward more deliberate, structure-based reasoning, consistent with a Dual-System account where removing ``assumption-inducing'' names reduces System~1 misfires and encourages System~2 tracing.

Overall, Python accuracy tends to increase with obfuscation, while JavaScript accuracy decreases. Two Python snippets (No.43, No.74) deviate, but comprehension remains relatively resilient: No.43 errors stay within the boolean domain (e.g., True/False swaps), and No.74 errors remain close to the ground truth \code{['hl', 'Hi']} (e.g., \code{['hi', 'hl']}), suggesting execution slips rather than major logic gaps. In contrast, the JavaScript outlier (No.136) shows higher accuracy under heavier obfuscation but deeper confusion: low-obfuscation errors cluster near \code{Null}, whereas high-obfuscation errors drift to unrelated outputs (e.g., \code{Infinity} or empty), indicating weak understanding despite accuracy~gain.

\begin{tcolorbox}[colback=white, colframe=black, arc=8pt, boxrule=0.5pt]
\textbf{RQ4 Takeaway:} {\em JavaScript follows a standard linear degradation: accuracy drops steadily as obfuscation increases ($L0 \text{ to } L3$), with control-flow distortion ($L2$) causing more harm than renaming ($L1$). In contrast, Python exhibits a counter-intuitive "inverse" trajectory. Accuracy increases under Renaming ($L1/L1b$) compared to Clean Code ($L0$), and L3 outperforms L2. This suggests that removing Python's semantic variable names forces users to abandon heuristics for careful logical tracing (System 2), while its strict formatting supports this deeper~reading. 
Conversely, JavaScript's flexible syntax appears to rely more heavily on variable names for clarity, leading to faster degradation when they are removed.}

\end{tcolorbox}
\section{Impact of programmer experience on performance (RQ5)}

\begin{figure}[t]
  \centering
  \begin{minipage}[t]{0.33\textwidth}
    \centering
    \includegraphics[width=\linewidth]{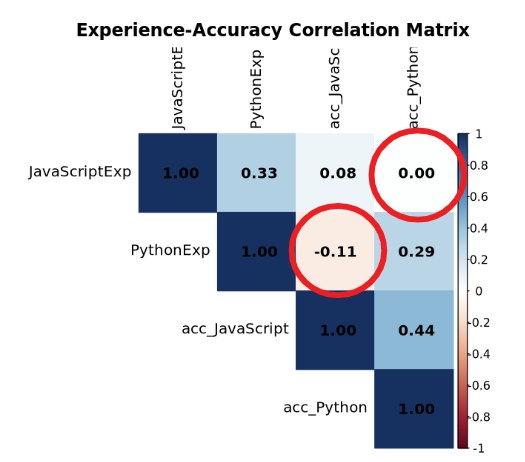}
    \vspace{-21pt}
    \caption{Experience-Acc. Correlation}
    \label{fig:correlation}
  \end{minipage}\hfill
  \begin{minipage}[t]{0.66\textwidth}
    \centering
    \includegraphics[width=\linewidth]{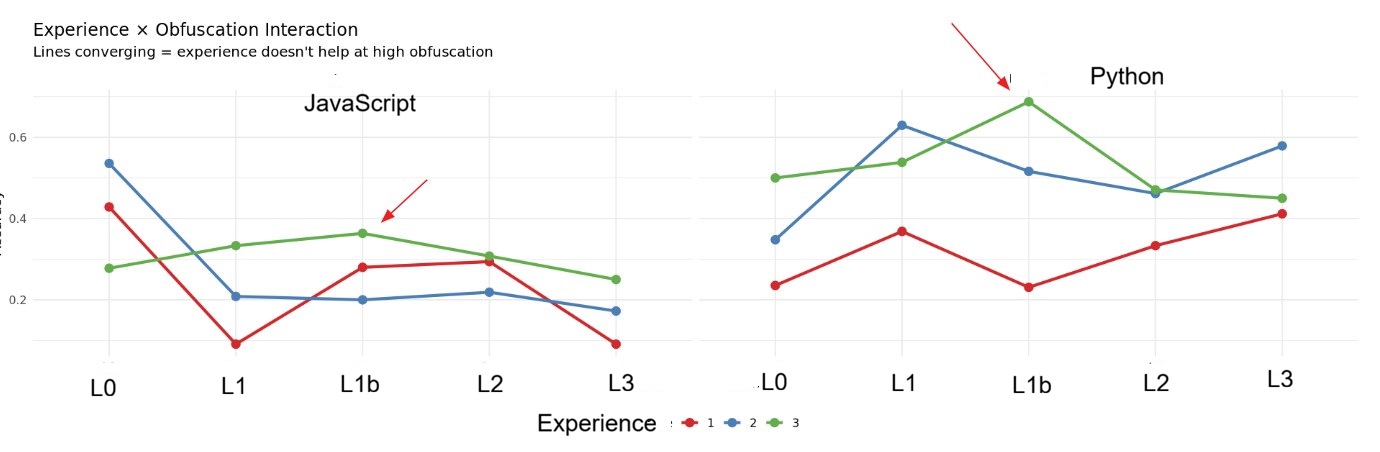}
    \vspace{-21pt}
    \caption{Experience and Accuracy by Obfuscation Levels}
    \label{fig:experience_obfuscation}
  \end{minipage}
\end{figure}

As seen in Fig.~\ref{fig:correlation}, the correlation matrix suggests a clear within-language relationship between experience and accuracy, alongside weak or even negative cross-language associations. Fig.~\ref{fig:correlation} and Table~\ref{fig:cross-language} show that experience predicts performance in Python (in a Pearson statistical significance test, $p$-value$\approx$0.01), where accuracy jumps from $0.321$ (no experience) to $0.508$ ($<1$ year). Conversely, JavaScript gains were not statistically significant ($0.250\to0.303, p=0.76$). This suggests Python tenure translates directly to task success, whereas JavaScript performance may depend more on general skill than language-specific experience. Moreover, while JavaScript experience does not predict Python accuracy, high Python experience incurs a steep penalty on JavaScript tasks. Furthermore, we see negative cross-language trends (e.g., Python experience $\to$ JavaScript accuracy dropping from $0.298$ to $0.244$). Python experts saw a 53.3\% relative accuracy decrease compared to novices (Bootstrap 95\% CI: $[-70.6\%, -31.6\%]$).  This suggests that language-specific proficiency can actively hinder performance in a new syntax, likely due to negative transfer of ingrained idioms and misfirings of System 1.

\begin{table}[t]
\centering
\caption{Contrast between In-Language skill acquisition and Cross-Language interference.}
\vspace{-12pt}
\label{tab:skill_transfer}
\footnotesize
\begin{tabular}{lcccc}
\toprule
& \multicolumn{2}{c}{\textbf{In-Language (Accuracy $\uparrow$)}} & \multicolumn{2}{c}{\textbf{Cross-Language (Accuracy $\downarrow$)}} \\
\cmidrule(lr){2-3} \cmidrule(lr){4-5}
\textbf{Exp. Level} & \textbf{Py Exp $\to$ Py Task} & \textbf{JS Exp $\to$ JS Task} & \textbf{Py Exp $\to$ JS Task} & \textbf{JS Exp $\to$ Py Task} \\
\midrule
None         & 0.321 & 0.250 & 0.298 & 0.469 \\
$\le 1$ year & 0.508 & 0.268 & 0.270 & 0.457 \\
$> 1$ year   & \textbf{0.522} & \textbf{0.303} & \textbf{0.244} & \textbf{0.455} \\
\bottomrule
\label{fig:cross-language}

\end{tabular}
\end{table}

Interestingly, this interference coexists with a "general aptitude" effect. JavaScript experience is unrelated to Python accuracy, and Python experience is slightly negatively related to JavaScript accuracy ($r = -0.11$). Accuracy in JavaScript and Python correlates positively ($r = 0.44$). Although Python tasks were statistically easier with 86\% higher odds of correctness ($Odds Ratio=2.31$), this correlation indicates that while years of experience drives interference, general tracing skill remains a transferable asset. Participants who perform well in one language generally excel in the other, likely due to broader skills such as control-flow tracing and verification that transcend~syntax.

\begin{wraptable}{r}{0.47\textwidth}
\centering
\caption{Accuracy by self-reported experience.}
\label{tab:joint_exp_perf}
\small
\vspace{-12pt}
\begin{tabular}{lccc}
\toprule
\textbf{JavaScript Exp.} & \multicolumn{3}{c}{\textbf{Python Experience}} \\
\cmidrule(lr){2-4}
 & None & $<1$ year & $\geq$1 year \\
\midrule
None         & 36.67 & 36.67 & 25.00 \\
$<1$ year    & 25.00 & 41.67 & 42.71 \\
$\geq$1 year & 50.00 & 39.58 & 34.72 \\
\bottomrule
\end{tabular}
\end{wraptable}

As shown in Table~\ref{tab:joint_exp_perf}, accuracy peaks in "mixed-profile" groups rather than among those most experienced in both languages. The strongest performance is concentrated among participants with moderate JavaScript experience ($\le1$ year) combined with any Python experience ($0.417$--$0.427$), as well as those with high JavaScript experience ($>1$ year) but no Python experience ($0.50$), though we interpret this specific cell cautiously due to low sample size ($n=1$). This latter case suggests that general programming skill can sometimes offset a lack of language tenure. Conversely, the lowest accuracies ($0.25$) occur when high experience in a language meets a total lack of experience in the other. This indicates a "bottleneck" effect, where a lack of basic syntax fluency hampers performance regardless of how familiar the other language is.

Notably, being "highly experienced in both" languages does not guarantee peak performance ($0.347$). In Dual-System terms, this paradox may reflect a strategy effect: while moderately experienced participants are forced to verify code more consistently, highly experienced participants may rely too heavily on fast, pattern-based System 1 expectations. Under the obfuscated conditions of~these tasks, such shortcuts can misfire, leading to errors that a more deliberate System 2 approach might avoid. Peak performance thus requires a balance between enough fluency for quick parsing and enough "friction" to trigger analytical verification when cues are unreliable.

A granular view of these trends in Table~\ref{fig:cross-language}, shows that in (Py Exp $\rightarrow$ Py Task), accuracy rises sharply from 0.321 (No Exp) to 0.508 (<=1 year) and 0.522 (>1 year).  The gain from <=1 year to >1 year is smaller than the jump from No Exp to <=1 year, implying that once participants have any Python experience, they gain a large boost, with only a small additional gain beyond one year.

\textbf{ Stratifying of Experience and Accuracy by Obfuscation Levels}.
As seen in Fig.~\ref{fig:experience_obfuscation}, in JavaScript, experience helps mainly with low/no obfuscation, and the advantage shrinks under heavier obfuscation. At L0, higher-experience groups are more accurate, but from L1--L3 the lines compress and often cross, indicating experience is not a stable predictor. Accuracy is low for all groups at L1 and L3, suggesting renaming and the strongest combined obfuscation disrupt comprehension enough that even experienced readers cannot reliably use usual shortcuts. This convergence is consistent with obfuscation suppressing the benefits of language fluency by degrading the cues experts typically exploit, pushing everyone toward similarly error-prone, effortful~reasoning.

\begin{wrapfigure}{r}{3.5in}
    \centering
    \includegraphics[width=0.65\textwidth]{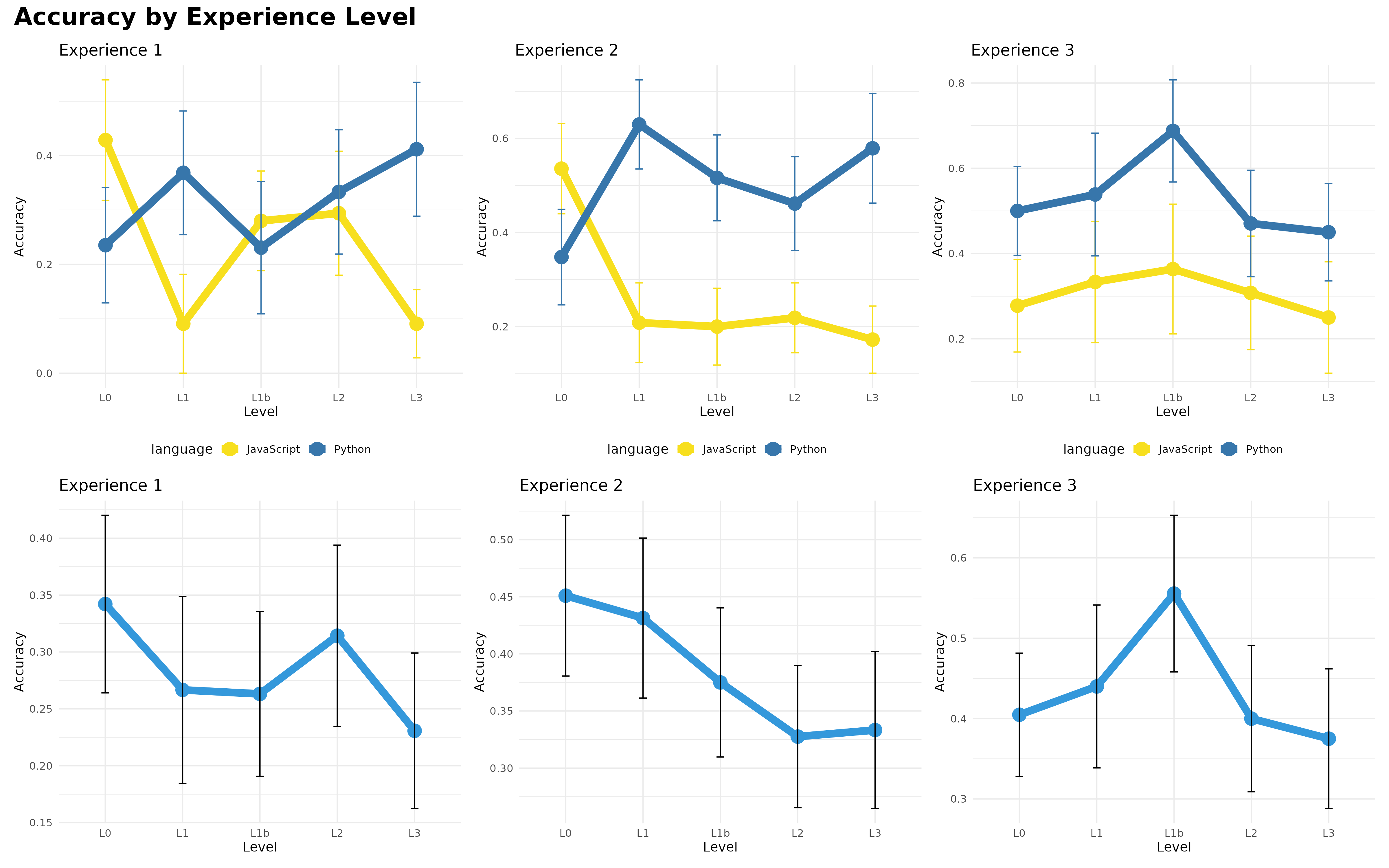}
    \vspace{-21pt}
    \caption{(Top) Accuracy by Experience, split by Language. (Bottom) Accuracy by Experience, overall.}
    \label{fig:lb1-experience}
\end{wrapfigure}

For Python, experience generally helps, but the benefit is tier-dependent and not monotonic. More-experienced groups usually outperform the least-experienced group, showing a clearer advantage than in JavaScript, but the gap varies: it is largest around L1b (adversarial renaming), where misleading names appear to punish novices and experience helps readers rely on deeper cues (e.g., data flow over identifiers). At higher obfuscation (L2/L3), the lines converge, suggesting diminishing returns once control flow is altered and everyone is forced into heavy System~2 tracing.
A useful interpretation is that 
when the task forces heavy System 2 (at higher obfuscation), the advantage of experience can wash out, producing converging lines and even occasional reversals. In contrast, JavaScript is more brittle overall, with low accuracy across obfuscated tiers for all experience groups, implying obfuscation more effectively neutralizes language-specific expertise.

\textbf{Accuracy by Experience}. As seen in Fig.~\ref{fig:lb1-experience}, the most experienced participants ($>1$year) show a distinct performance peak at L1b (adversarial renaming) across both languages. In Dual-System terms, this mismatch prompts a shift from fast System 1 heuristics toward earlier System 2 verification, allowing experts to prioritize deeper cues like data/control dependencies over identifier semantics. However, at the heaviest obfuscation (L3), accuracy remains low for all groups, suggesting that the complexity of combined obfuscations can overwhelm even deliberate System 2 reasoning.

\begin{wrapfigure}{r}{0.40\textwidth}
    \centering
    \includegraphics[width=0.40\textwidth]{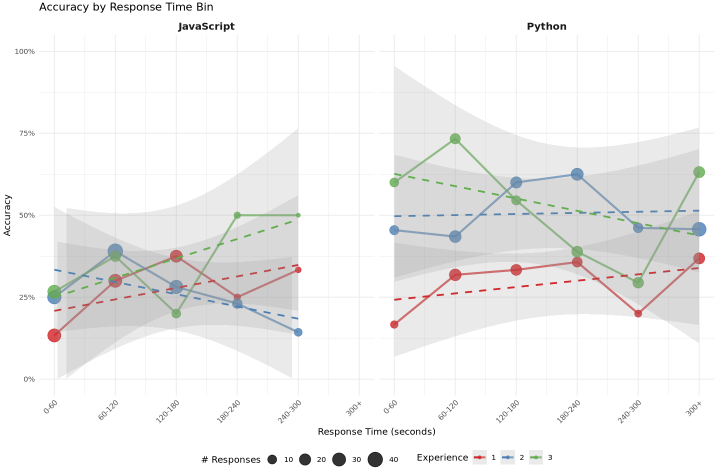}
    \vspace{-14pt}
    \caption{Response Time vs. Accuracy for different experience levels and divided by sixty second bins.}
    \label{fig:accuracy_by_response_bin}
\end{wrapfigure}

Python is {\em consistently higher than JavaScript at most experience levels and tiers.} In the top row, Python sits above JavaScript for nearly all obfuscation levels in Experience 2 and 3, indicating stronger robustness on Python. JavaScript {\em is especially sensitive to renaming (L1/L3) for less experienced participants.} In Experience 1, JavaScript drops sharply from L0 to L1 and is low again at L3, consistent with novices' name-driven System 1 heuristics being disrupted by identifier changes.

Experience {\em reduces the ''cost'' of obfuscation more for Python than for JavaScript.} From Experience 1 to 3, Python improves and stabilizes across tiers, whereas JavaScript stays comparatively low and more volatile under obfuscation--especially at L3.

Error bars {\em widen under harder tiers and lower experience.} Uncertainty is larger (for JavaScript in harder tiers), indicating greater between-subject variability that shrinks with experience in Python.

\textbf {Accuracy by Response Time for Different Experience Levels}.
Fig.~\ref{fig:accuracy_by_response_bin} shows a non-linear time-accuracy relationship that varies by experience and language. For Experience 1 and 2 (red/blue), accuracy follows an inverted-U: it rises from very fast responses (0-60s) to mid-range bins (roughly 60-240s, depending on language), then drops at the longest times (240-300s+). This fits a Dual-System view: fast answers reflect error-prone System 1 heuristics, mid-range times reflect productive System 2 tracing, and very long times increasingly signal confusion.

For Experience 3 (green), the pattern differs across both languages. Accuracy peaks early (60-120s), dips in the middle bins, then rebounds at the slowest end in Python (a bump at 300+s) and also in JavaScript. This suggests long times for experts may reflect via System 2 verification rather than being~stuck.

Across bins, Python accuracy is higher than JavaScript at the same experience level (Experience 2--3). Bubble sizes shrink at the extremes, so tail patterns are noisier, but moderate time helps less experienced subjects most, while experts can turn long deliberation into recovered~accuracy.

\begin{tcolorbox}[colback=white, colframe=black, arc=8pt, boxrule=0.5pt]
    \textbf{RQ5 Takeaway:} \textit{Linear correlation within-language is strong but cross-language correlation is negative. Experience is not purely additive across languages: outcomes depend on the combination of fluency and reasoning strategy, with several ``mixed'' experience profiles outperforming the ``max experience'' profile.  Experience helps most when the obfuscation attacks misleading surface cues (L1b) and less when obfuscation disrupts program structure (L3).}
\end{tcolorbox}

\section{Key Findings}

{\bf Obfuscation forces a shift from heuristic to deliberate processing.} The presence of obfuscation (L1--L3) triggered an immediate spike in response time and a drop in accuracy, signaling a transition from fast System 1 heuristics to effortful System 2 tracing. While deliberation generally improved accuracy, very long response times often correlated with incorrectness, indicating a breakdown of the mental model. This shift is clearest at L1b (adversarial renaming), which functioned as a "trap" for System 1; misleading-but-plausible names baited fast, intuitive errors, requiring significantly longer deliberation for humans to consciously verify and correct initial interpretations.

{\bf Obfuscation does not imply a linear drop in accuracy.} 
More obfuscation does not always equals lower accuracy. The relation is non-monotonic and language-dependent. JavaScript followed the expected linear decline, lowest accuracy with the hardest tier (L3). Python shows paradoxical results as renaming (L1) sometimes outperformed the baseline (L0). This suggests that code understanding is driven by interactions between language-specific cues (syntax, idioms) and~obfuscation. 

{\bf Experience Transfer.}
Proficiency helps, but largely within-language. Experience correlates with accuracy within the same language but transfers weakly across languages under obfuscation.  At the same time, accuracy across languages correlates positively, hinting at a general code comprehension/tracing skill that is distinct from language proficiency. 

{\bf Code Complexity}. Code understandability here is driven by factors other than
variable name length. Control-flow obfuscation becomes increasingly harmful as programs get longer, whereas pure renaming can sometimes be mitigated--or even partially offset--by longer, richer structure that provides more traceable cues. For each CC level (low, medium, high), the accuracy is non-monotonic with respect to the increasing obfuscation levels but rewarded checking and System 2 thinking.

\section{Implications}

{\bf \underline{Implications for program-comprehension and security researchers.}} 

{\bf Reconsider the ``monotonic difficulty'' assumption}. The observed non-monotonicity--particu\-larly the Python renaming gains--suggests obfuscation can sometimes shift strategy rather than purely degrade comprehension. Future work should evaluate obfuscation with multiple behavioral signals (accuracy, response-time distributions, fast/slow splits), not solely aggregate correctness.

{\bf Cognitive Complexity != Computational Complexity}. Traditional metrics like cyclomatic complexity and identifier length failed to predict human effort in our study. Future security evaluations must move beyond static metrics to model 'cognitive load', specifically measuring the disruption of System 1 heuristics versus the taxation of System 2 tracing.

{\bf Model obfuscation as a two-part cognitive intervention.} Results are consistent with a Dual-System framing: obfuscation (i) disrupts System-1 shortcuts (fast answers fail) and (ii) taxes System-2 tracing (time increases, with eventual confusion at extremes). This perspective supports more precise hypotheses about when humans fail (heuristic traps vs. tracing overload) and motivates finer-grained designs (e.g., manipulating misleadingness vs. informativeness of names).

{\bf Treat adversarial identifiers as a first-class experimental factor.}
L1b's signature suggests ``identifier semantics'' is not merely noise: misleading names can actively distort mental models. This opens a path for systematic measures of semantic obfuscation (e.g., contextual mismatch scores) and for benchmarks designed to distinguish ``uninformative'' from ``misleading''~obfuscation.

{\bf \underline{Implications for practitioners using obfuscation}}

{\bf Control-flow transformations are the most reliably damaging lever, but effectiveness is context-dependent.}
Control-flow alterations tend to produce the largest comprehension costs and confusion patterns. However, language- and population-specific effects matter: sometime, renaming may not reduce comprehension as much as expected and can even prompt more careful~reading.

{\bf Obfuscation strategy should be language- and threat-model specific.}
If the adversary is a human analyst, the ``best'' obfuscation may differ between Python and JavaScript, and between novices and experts. One should not assume that a obfuscation recipe generalizes across ecosystems.

{\bf Adversarial renaming} may seem like cheap ways to obfuscate code, but their effectiveness is language and model specific. In certain cases, renaming may help humans focus. 

{\bf \underline{Implications for tool builders}}

{\bf Obfuscation and deobfuscation tools should incorporate human-centric metrics}.
Static complexity or functional equivalence is not enough to characterize human difficulty. Tools could report expected impacts on (i) heuristic reliability (probability of fast failure) and (ii) tracing burden (time increases and confusion risk), enabling more informed choices about transformations.

{\bf Deobfuscation UX should target the bottleneck introduced by each tier}.
For control-flow obfuscation, restoring structural readability (e.g., simplified dispatch patterns, reconstructed blocks) may yield the biggest comprehension gains. For L1b, assistance should emphasize skepticism on semantics and provide scaffolds for dataflow verification rather than simply meaningful names.

{\bf Experience-aware assistance should be language-aware.}
Since experience transfers imperfectly, assistants should not overgeneralize user proficiency. They should detect conditions where System-1 expectations are likely to misfire (e.g., L1b) and nudge users toward targeted System-2.

\vspace{-6pt}
\section{Related Work}

Ceccato et al.~\cite{ceccato2014family} measure how obfuscation affects human time and task success. They show that simpler techniques (e.g., identifier renaming) can be more effective than more complex ones (e.g., opaque predicates). Their earlier work examined how obfuscation reduces attackers' efficiency~\cite{ceccato2009effectiveness}. From Vitticchie {\em et al.}~\cite{viticchie2016assessment},  obfuscation has significant effect on time-to-complete and successful attack efficiency. Regano {\em et al.}~\cite{regano2025empiricalassessmentcodecomprehension}
evaluate the effect of layering obfuscation techniques and the correlation between objective metrics of the attacked code and the likelihood of a successful attack. Wang {\em et al.}~\cite{wang2018softwareprotection} studied the deployment of obfuscation techniques in mobile software development.

Several researchers have studied the impact of naming to code understanding~\cite{BuseWeimer2008MetricReadability,DBLP:journals/tse/BuseW10,AvidanFeitelson2017VariableNames,LawrieMorrellFeildBinkley2006WhatsInAName,LawrieMorrellFeildBinkley2007EffectiveIdentifierNames,LawrieFeildBinkley2007AbbrevIdentifiers,HofmeisterSiegmundHolt2019ShorterNames,BoerstlerPaech2016MethodChainsComments,Eom2024R2I}.
Rajlich and Wilde argue that developers understand code by mapping it to high-level concepts~\cite{rajlich2002role}. Many proposed proxies, cyclomatic complexity, Halstead volume, Cognitive Complexity~\cite{mccabe_cc,halstead1977elements,cognitivecomplexity2017}--show weak empirical support: Scalabrino \textit{et al.} tested 121 metrics and found none strongly correlated with perceived or actual understandability~\cite{scalabrino2021automatically}. Trockman \textit{et al.} achieve only modest gains by combining metrics with statistical models~\cite{Trockman2018}. Readability models predict surface style judgments, but readability reflects clarity more than true comprehension.

Eye-tracking studies found novices read code linearly, like texts, while experts use non-linear strategies~\cite{Crosby1990}. Zhang {\em et al.}~\cite{siegmund2014understanding} merge human and model attention for code summarization. The authors later mimick human visual attention to improve code LLMs~\cite{zhang2025eyemulatorimprovingcodelanguage}.
Other researchers~\cite{siegmund2014understanding,siegmund2017measuring,Busjahn2015} further showed experts exhibit more non-linear gaze patterns than novices' sequential reading.

\section{Threats to Validity}

\textbf{External Validity.} Our student participants may differ in skill from professional reverse engineers. Our tasks use short, self-contained HumanEval-X functions, enabling controlled measurement but not reflecting large-scale systems. Generalizability is limited by our language choices (Python/JS) and by omitting advanced transformations (e.g., opaque predicates, data obfuscation, virtualization).

\textbf{Internal Validity.} The 75-minute time limit could shift strategies (e.g., rushing later items). Different obfuscation tools for language may add implementation artifacts beyond the intended changes. Presenting a question to less experienced subject may introduce learning or fatigue effects. 

\textbf{Construct Validity.} Time-to-completion is a proxy for cognitive effort: it is confounded by reading speed/strategy, and short times may reflect either expertise or guessing. Qualtrics timers capture only end-to-end duration (incl. interface overhead) and miss fine-grained behaviors (pauses, rereading). To reduce the off-task delays, we had three proctors in the room to remind the students.
Self-reported experience is subjective. Our tiers reflect real obfuscation families, but their instantiations depend on tool details. L1-L3 should be read as representative rather than~canonical.

\balance

\bibliographystyle{ACM-Reference-Format}

\bibliography{references}

@ARTICLE{scalabrino2021automatically,
  author={Scalabrino, Simone and Bavota, Gabriele and Vendome, Christopher and Linares-V{\'a}squez, Mario and Poshyvanyk, Denys and Oliveto, Rocco},
  journal={IEEE Transactions on Software Engineering}, 
  title={Automatically Assessing Code Understandability}, 
  year={2021},
  volume={47},
  number={3},
  pages={595-613},
  doi={10.1109/TSE.2019.2901468}}

@inproceedings{rajlich2002role,
author = {Rajlich, V\'{a}clav and Wilde, Norman},
title = {The Role of Concepts in Program Comprehension},
year = {2002},
isbn = {0769514952},
publisher = {IEEE Computer Society},
address = {USA},
booktitle = {Proceedings of the 10th International Workshop on Program Comprehension},
pages = {271},
series = {IWPC '02}
}

@inproceedings{Trockman2018,
author = {Asher Trockman and Keenen Cates and Mark Mozina and Tuan Nguyen and Christian K\"{a}stner and Bogdan Vasilescu},
title = {{\textquotedblleft}Automatically Assessing Code Understandability{\textquotedblright} Reanalyzed: Combined Metrics Matter},
booktitle = {Proc. 15th Int. Conf. Mining Software Repositories (MSR)},
pages = {46--57},
year = {2018},
publisher = {ACM},
doi = {10.1145/3196398.3196441},
url = {https://doi.org/10.1145/3196398.3196441}
}

@article{Crosby1990,
author = {Martha E. Crosby and Jan Stelovsky},
title = {How do we read algorithms? A case study},
journal = {{IEEE} Computer},
volume = {23},
number = {1},
pages = {25--35},
year = {1990},
doi={10.1109/2.48797}}

@inproceedings{Busjahn2015,
author = {Teresa Busjahn and Roman Bednarik and Andrew Begel and Martha Crosby and James Paterson and Carsten Schulte and Bonita Sharif and Sascha Tamm},
title = {Eye Movements in Code Reading: Relaxing the Linear Order},
booktitle = {Proc. 23rd {IEEE} Int. Conf. Program Comprehension (ICPC)},
pages = {255--265},
year = {2015},
publisher = {IEEE},
doi = {10.1109/ICPC.2015.36}
}

@book{halstead1977elements,
  title={Elements of Software Science (Operating and Programming Systems Series)},
  author={Halstead, Maurice H.},
  year={1977},
  publisher={Elsevier Science Inc.}
}

@inproceedings{cognitivecomplexity2017,
author = {Campbell, G. Ann},
title = {Cognitive complexity: an overview and evaluation},
year = {2018},
isbn = {9781450357135},
publisher = {Association for Computing Machinery},
address = {New York, NY, USA},
url = {https://doi.org/10.1145/3194164.3194186},
doi = {10.1145/3194164.3194186},
booktitle = {Proceedings of the 2018 International Conference on Technical Debt},
pages = {57--58},
numpages = {2},
location = {Gothenburg, Sweden},
series = {TechDebt '18}
}

@techreport{collberg1997taxonomy,
    author = {Collberg, Christian and Thomborson, C. and Low, Douglas},
    title = {A Taxonomy of Obfuscating Transformations},
    institution = {University of Auckland},
    year = {1997},
    month = {07},
    number = {148},
    url = {https://researchspace.auckland.ac.nz/handle/2292/3491}
}

@inproceedings{ceccato2008towards,
author = {Ceccato, Mariano and Di Penta, Massimiliano and Nagra, Jasvir and Falcarin, Paolo and Ricca, Filippo and Torchiano, Marco and Tonella, Paolo},
title = {Towards experimental evaluation of code obfuscation techniques},
year = {2008},
isbn = {9781605583211},
publisher = {Association for Computing Machinery},
address = {New York, NY, USA},
url = {https://doi.org/10.1145/1456362.1456371},
doi = {10.1145/1456362.1456371},
booktitle = {Proceedings of the 4th ACM Workshop on Quality of Protection},
pages = {39--46},
numpages = {8},
location = {Alexandria, Virginia, USA},
series = {QoP '08}
}

@inproceedings{laszlo2007cff,
author = {L{\'a}szl{\'o}, T{\'i}mea and Kiss, {\'A}kos},
year = {2007},
month = {06},
pages = {},
title = {Obfuscating C++ Programs via Control Flow Flattening},
volume = {30},
journal = {Annales Universitatis Scientiarum Budapestinensis de Rolando E{\"o}tv{\"o}s Nominatae. Sectio Computatorica}
}

@article{evans2013dual,
author = {Jonathan St. B. T. Evans and Keith E. Stanovich},
title ={Dual-Process Theories of Higher Cognition: Advancing the Debate},

journal = {Perspectives on Psychological Science},
volume = {8},
number = {3},
pages = {223-241},
year = {2013},
doi = {10.1177/1745691612460685},
    note ={PMID: 26172965},
URL = {https://doi.org/10.1177/1745691612460685}
}

@inproceedings{Schulte2008BlockModel,
  author    = {Carsten Schulte},
  title     = {Block Model: an educational model of program comprehension as a tool for a scholarly approach to teaching},
  booktitle = {Proceedings of the Fourth International Workshop on Computing Education Research (ICER '08)},
  pages     = {149--160},
  publisher = {ACM},
  year      = {2008},
  doi       = {10.1145/1404520.1404535}
}

@INPROCEEDINGS{viticchie2016assessment,
  author={Viticchi{\'e}, Alessio and Regano, Leonardo and Torchiano, Marco and Basile, Cataldo and Ceccato, Mariano and Tonella, Paolo and Tiella, Roberto},
  booktitle={2016 IEEE 16th International Working Conference on Source Code Analysis and Manipulation (SCAM)}, 
  title={Assessment of Source Code Obfuscation Techniques}, 
  year={2016},
  volume={},
  number={},
  pages={11-20},
  doi={10.1109/SCAM.2016.17}}

@inproceedings{siegmund2014understanding,
author = {Siegmund, Janet and K\"{a}stner, Christian and Apel, Sven and Parnin, Chris and Bethmann, Anja and Leich, Thomas and Saake, Gunter and Brechmann, Andr\'{e}},
title = {Understanding understanding source code with functional magnetic resonance imaging},
year = {2014},
isbn = {9781450327565},
publisher = {Association for Computing Machinery},
address = {New York, NY, USA},
url = {https://doi.org/10.1145/2568225.2568252},
doi = {10.1145/2568225.2568252},
booktitle = {Proceedings of the 36th International Conference on Software Engineering},
pages = {378--389},
numpages = {12},
location = {Hyderabad, India},
series = {ICSE 2014}
}

@inproceedings{siegmund2017measuring,
author = {Siegmund, Janet and Peitek, Norman and Parnin, Chris and Apel, Sven and Hofmeister, Johannes and K\"{a}stner, Christian and Begel, Andrew and Bethmann, Anja and Brechmann, Andr\'{e}},
title = {Measuring neural efficiency of program comprehension},
year = {2017},
isbn = {9781450351058},
publisher = {Association for Computing Machinery},
address = {New York, NY, USA},
url = {https://doi.org/10.1145/3106237.3106268},
doi = {10.1145/3106237.3106268},
booktitle = {Proceedings of the 2017 11th Joint Meeting on Foundations of Software Engineering},
pages = {140--150},
numpages = {11},
location = {Paderborn, Germany},
series = {ESEC/FSE 2017}
}

@article{Stroop1935Interference,
  author  = {Stroop, J. Ridley},
  title   = {Studies of interference in serial verbal reactions},
  journal = {Journal of Experimental Psychology},
  year    = {1935},
  volume  = {18},
  number  = {6},
  pages   = {643--662},
  doi     = {10.1037/h0054651},
  url     = {https://doi.org/10.1037/h0054651}
}

@inproceedings{zheng2023codegeex,
  title={CodeGeeX: A Pre-Trained Model for Code Generation with Multilingual Benchmarking on HumanEval-X},
  author={Qinkai Zheng and Xiao Xia and Xu Zou and Yuxiao Dong and Shan Wang and Yufei Xue and Zihan Wang and Lei Shen and Andi Wang and Yang Li and Teng Su and Zhilin Yang and Jie Tang},
  booktitle={Proceedings of the 29th ACM SIGKDD Conference on Knowledge Discovery and Data Mining},
  pages={5673--5684},
  year={2023},
  doi = {10.1145/3580305.3599790},
  url = {https://doi.org/10.1145/3580305.3599790}
}

@article{husain2019codesearchnet, 
    title={{CodeSearchNet} challenge: Evaluating the state of semantic code search}, 
    author={Husain, Hamel and Wu, Ho-Hsiang and Gazit, Tiferet and Allamanis, Miltiadis and Brockschmidt, Marc}, 
    journal={arXiv preprint arXiv:1909.09436},
    url={https://arxiv.org/abs/1909.09436},
    year={2019}
}

@software{obfuXtreme25,
  author       = {spyboy-productions},
  title        = {ObfuXtreme},
  url          = {https://github.com/spyboy-productions/ObfuXtreme},
  note         = {GitHub repository, last accessed September 24, 2025}
}

@software{javascript_obfuscator_411,
  author       = {{javascript-obfuscator contributors}},
  title        = {javascript-obfuscator},
  version      = {4.1.1},
  year         = {2025},
  url          = {https://github.com/javascript-obfuscator/javascript-obfuscator},
  note         = {JavaScript obfuscation tool, package version 4.1.1, last accessed October 17, 2025}
}

@article{10.1145/2886012,
author = {Schrittwieser, Sebastian and Katzenbeisser, Stefan and Kinder, Johannes and Merzdovnik, Georg and Weippl, Edgar},
title = {Protecting Software through Obfuscation: Can It Keep Pace with Progress in Code Analysis?},
year = {2016},
issue_date = {March 2017},
publisher = {Association for Computing Machinery},
address = {New York, NY, USA},
volume = {49},
number = {1},
issn = {0360-0300},
url = {https://doi.org/10.1145/2886012},
doi = {10.1145/2886012},
journal = {ACM Comput. Surv.},
month = apr,
articleno = {4},
numpages = {37}
}

@ARTICLE{mccabe_cc,
  author={McCabe, T.J.},
  journal={IEEE Transactions on Software Engineering}, 
  title={A Complexity Measure}, 
  year={1976},
  volume={SE-2},
  number={4},
  pages={308-320},
  doi={10.1109/TSE.1976.233837}}

@INPROCEEDINGS{ceccato2009effectiveness,
  author={Ceccato, Mariano and Di Penta, Massimiliano and Nagra, Jasvir and Falcarin, Paolo and Ricca, Filippo and Torchiano, Marco and Tonella, Paolo},
  booktitle={2009 IEEE 17th International Conference on Program Comprehension}, 
  title={The effectiveness of source code obfuscation: An experimental assessment}, 
  year={2009},
  volume={},
  number={},
  pages={178-187},
  doi={10.1109/ICPC.2009.5090041}}

@article{ceccato2014family,
author = {Ceccato, Mariano and Penta, Massimiliano and Falcarin, Paolo and Ricca, Filippo and Torchiano, Marco and Tonella, Paolo},
title = {A family of experiments to assess the effectiveness and efficiency of source code obfuscation techniques},
year = {2014},
issue_date = {August 2014},
publisher = {Kluwer Academic Publishers},
address = {USA},
volume = {19},
number = {4},
issn = {1382-3256},
url = {https://doi.org/10.1007/s10664-013-9248-x},
doi = {10.1007/s10664-013-9248-x},
journal = {Empirical Softw. Engg.},
month = aug,
pages = {1040--1074},
numpages = {35}
}

@misc{regano2025empiricalassessmentcodecomprehension,
      title={Empirical Assessment of the Code Comprehension Effort Needed to Attack Programs Protected with Obfuscation}, 
      author={Leonardo Regano and Daniele Canavese and Cataldo Basile and Marco Torchiano},
      year={2025},
      eprint={2511.21301},
      archivePrefix={arXiv},
      primaryClass={cs.CR},
      url={https://arxiv.org/abs/2511.21301}, 
}

@inproceedings{wang2018softwareprotection,
author = {Wang, Pei and Bao, Qinkun and Wang, Li and Wang, Shuai and Chen, Zhaofeng and Wei, Tao and Wu, Dinghao},
title = {Software protection on the go: a large-scale empirical study on mobile app obfuscation},
year = {2018},
isbn = {9781450356381},
publisher = {Association for Computing Machinery},
address = {New York, NY, USA},
url = {https://doi.org/10.1145/3180155.3180169},
doi = {10.1145/3180155.3180169},
booktitle = {Proceedings of the 40th International Conference on Software Engineering},
pages = {26--36},
numpages = {11},
location = {Gothenburg, Sweden},
series = {ICSE '18}
}

@misc{zhang2025eyemulatorimprovingcodelanguage,
      title={EyeMulator: Improving Code Language Models by Mimicking Human Visual Attention}, 
      author={Yifan Zhang and Chen Huang and Yueke Zhang and Jiahao Zhang and Toby Jia-Jun Li and Collin McMillan and Kevin Leach and Yu Huang},
      year={2025},
      eprint={2508.16771},
      archivePrefix={arXiv},
      primaryClass={cs.SE},
      url={https://arxiv.org/abs/2508.16771}, 
}

@inproceedings{BuseWeimer2008MetricReadability,
  author    = {Raymond P. L. Buse and Westley Weimer},
  title     = {A Metric for Software Readability},
  booktitle = {Proceedings of the 2008 International Symposium on Software Testing and Analysis (ISSTA '08)},
  pages     = {121--130},
  year      = {2008},
  publisher = {ACM},
  doi       = {10.1145/1390630.1390647},
  url       = {https://doi.org/10.1145/1390630.1390647}
}

@article{DBLP:journals/tse/BuseW10,
  author  = {Raymond P. L. Buse and Westley Weimer},
  title   = {Learning a Metric for Code Readability},
  journal = {IEEE Trans. Software Eng.},
  volume  = {36},
  number  = {4},
  pages   = {546--558},
  year    = {2010},
  doi     = {10.1109/TSE.2009.70},
  url     = {https://doi.org/10.1109/TSE.2009.70}
}

@inproceedings{AvidanFeitelson2017VariableNames,
  author    = {Eran Avidan and Dror G. Feitelson},
  title     = {Effects of Variable Names on Comprehension: An Empirical Study},
  booktitle = {2017 IEEE/ACM 25th International Conference on Program Comprehension (ICPC)},
  pages     = {55--65},
  year      = {2017},
  doi       = {10.1109/ICPC.2017.27},
  url       = {https://doi.org/10.1109/ICPC.2017.27}
}

@inproceedings{LawrieMorrellFeildBinkley2006WhatsInAName,
  author    = {Dawn Lawrie and Christopher Morrell and Henry Feild and David Binkley},
  title     = {What's in a Name? A Study of Identifiers},
  booktitle = {14th IEEE International Conference on Program Comprehension (ICPC 2006)},
  pages     = {3--12},
  year      = {2006},
  doi       = {10.1109/ICPC.2006.51},
  url       = {https://doi.org/10.1109/ICPC.2006.51}
}

@article{LawrieMorrellFeildBinkley2007EffectiveIdentifierNames,
  author  = {Dawn J. Lawrie and Christopher Morrell and Henry Feild and David Binkley},
  title   = {Effective identifier names for comprehension and memory},
  journal = {Innovations in Systems and Software Engineering},
  year    = {2007},
  doi     = {10.1007/s11334-007-0031-2},
  url     = {https://doi.org/10.1007/s11334-007-0031-2}
}

@inproceedings{LawrieFeildBinkley2007AbbrevIdentifiers,
  author    = {Dawn Lawrie and Henry Feild and David Binkley},
  title     = {Extracting Meaning from Abbreviated Identifiers},
  booktitle = {Seventh IEEE International Working Conference on Source Code Analysis and Manipulation (SCAM 2007)},
  pages     = {213--222},
  year      = {2007},
  publisher = {IEEE Computer Society},
  doi={10.1109/SCAM.2007.17},
  url = {https://doi.org/10.1109/SCAM.2007.17}
}

@article{HofmeisterSiegmundHolt2019ShorterNames,
  author  = {Johannes C. Hofmeister and Janet Siegmund and Daniel V. Holt},
  title   = {Shorter identifier names take longer to comprehend},
  journal = {Empirical Software Engineering},
  year    = {2019},
  doi     = {10.1007/s10664-018-9621-x},
  url     = {https://doi.org/10.1007/s10664-018-9621-x}
}

@article{BoerstlerPaech2016MethodChainsComments,
  author  = {J{\"u}rgen B{\"o}rstler and Barbara Paech},
  title   = {The Role of Method Chains and Comments in Software Readability and Comprehension---An Experiment},
  journal = {IEEE Transactions on Software Engineering},
  volume  = {42},
  number  = {9},
  pages   = {886--898},
  year    = {2016},
  doi     = {10.1109/TSE.2016.2527791},
  url     = {https://doi.org/10.1109/TSE.2016.2527791}
}

@article{Eom2024R2I,
  author = {Eom, Haeun and Kim, Dohee and Lim, Sori and Koo, Hyungjoon and Hwang, Sungjae},
  title   = {R2I: A Relative Readability Metric for Decompiled Code},
  journal = {Proceedings of the ACM on Software Engineering},
  year    = {2024},
  doi     = {10.1145/3643744},
  url     = {https://doi.org/10.1145/3643744}
}

\end{document}